\documentclass{aastex63}
\usepackage{amsmath}

\usepackage{graphicx}
\usepackage{amsmath}
\usepackage{mathtools}
\usepackage{kantlipsum}
\usepackage{xparse}
\usepackage{enumerate}
\usepackage[version=3]{mhchem}
\usepackage{amsmath,array}
\usepackage[T1]{fontenc}
\usepackage{booktabs}

\usepackage{natbib}

\setcitestyle{super}

\usepackage{sidecap}
\usepackage{tabularx}

\graphicspath{{./}{figures/}}

\begin{document}
\renewcommand{\abstractname}{Introduction}
\title{Helium Enhanced Planets Along the Upper Edge of the Radius Valley}

\correspondingauthor{Isaac Malsky}
\email{imalsky@umich.edu}

\author[0000-0003-0217-3880]{Isaac Malsky}
\affil{Department of Astronomy and Astrophysics, University of Michigan, Ann Arbor, MI, 48109, USA}

\author[0000-0003-0638-3455]{Leslie Rogers}
\affil{Department of Astronomy and Astrophysics, University of Chicago, Chicago IL, 60637, USA}

\author[0000-0002-1337-9051]{Eliza M.-R.\ Kempton}
\affil{Department of Astronomy, University of Maryland, College Park, MD 20742, USA}

\author[0000-0002-1197-6143]{Nadejda Marounina}
\affil{Department of Astronomy and Astrophysics, University of Chicago, Chicago IL, 60637, USA}


\begin{abstract}
The low mean densities of sub-Neptunes imply that they formed within a few million years and accreted primordial envelopes\cite{Rogers2015ApJ}. Because these planets receive a total X-ray and extreme ultra-violet flux that is comparable to the gravitational binding energy of their envelopes, their primordial hydrogen-helium atmospheres are susceptible to mass loss\cite{Owen&Wu2013ApJ}. Models of photoevaporating sub-Neptunes have so far assumed that envelope compositions remain constant over time. However, preferential loss of atmospheric hydrogen has the potential to change their compositions. Here, by modeling the thermal and compositional evolution of sub-Neptunes undergoing atmospheric escape with diffusive separation between hydrogen and helium, we show that planets with radii between 1.6 and 2.5 R$_\oplus$ can become helium-enhanced from billions of years of photoevaporation, obtaining helium mass fractions in excess of 40\%. Atmospheric helium enhancement can be detected through transmission spectra, providing a novel observational test for whether atmospheric escape creates the radius valley\cite{FultonEt2017AJ}.
\end{abstract}

\section*{}
Planets with orbital periods shorter than 100 days and radii smaller than Neptune outnumber planets larger than Neptune by a factor of ten\cite{Fulton&Petigura2018AJ}. The {\it Kepler} survey has revealed that the radius distribution of this population is bimodal: there is a scarcity of planets between 1.5 and 2.0 $R_{\oplus}$ and peaks in the occurrence rate at $\sim$1.3~$R_{\oplus}$ (super-Earths) and $\sim$2.4~$R_{\oplus}$ (sub-Neptunes)\cite{FultonEt2017AJ}. Planet mass-radius measurements show that most planets larger than 1.6 R$_{\oplus}$ have low mean densities, requiring voluminous volatile envelopes\cite{Rogers2015ApJ}. Although exoplanet surveys have begun to constrain the structure of this population, much of the formation and compositional evolution of sub-Neptunes remain a mystery, and a large variety of bulk compositions are \textit{a priori} possible\cite{Rogers&Seager2010bApJ}. Primordial hydrogen and helium accreted from the protoplanetary disk, H$_2$O accreted in the form of solid icy material, and atmospheric H$_2$O created through magma-atmosphere interactions and volcanic outgassing\cite{2021ApJ...909L..22K} may all contribute to the volatile envelopes of sub-Neptunes.

The radius valley could be explained as the outcome of multiple planet formation pathways. Atmospheric escape may shape the evolution of highly irradiated sub-Neptunes, bifurcating the population based on envelope retention\cite{2017ApJ...847...29O}. The primordial hydrogen-helium envelopes surrounding sub-Neptunes are susceptible to mass loss driven by ionizing radiation from their host star\cite{1981Icar...48..150W} and the thermal energy released from the planetary cores\cite{2018MNRAS.476..759G}. Planets that retain their envelopes may comprise the larger 2.4 R$_{\oplus}$ sub-Neptune mode of the radius distribution, while the evaporated cores of former sub-Neptunes may make up the smaller $\sim$ 1.3 R$_{\oplus}$ mode of the distribution.

Alternatively, if a significant sub-population of planets formed at or beyond the water snow line in the nascent protoplanetary disk before migrating inwards toward the star, the bimodality of the small planet radius distribution would reflect the differing bulk compositions of these two populations \cite{ZengEt2019PNAS}. In this scenario, while the small super-Earth mode of the planet radius distribution would still be comprised of rocky planets formed {\it in situ}, the larger sub-Neptune mode would be comprised of water-rich planets with several tens of percent water by mass \cite{waterworlds2003, water2004}.

Tests of the origin of the radius valley have so far relied on characterizing the radius distribution of planets as a function of characteristics such as orbital period, host star spectral type, stellar age, and stellar metallicity\cite{2015MNRAS.452.2127S, FultonEt2017AJ, 2020AJ....160..108B, Petigura2022}. Here we propose a new indicator of how the radius valley is sculpted by the photoevaporation of primordial envelopes: measuring the imprint of fractionated mass-loss on the atmospheric compositions of planets along the upper edge of the radius valley. To date, most models of the evolution of sub-Neptune-size planets experiencing atmospheric escape have assumed that the chemical composition of the planetary envelope remains constant over time as the envelope is gradually lost\cite{Owen&Wu2013ApJ,Lopez&Fortney2013ApJ,Chen&Rogers2016ApJ} . This approximation may be appropriate in the first $\sim 100$~Myr of a planet's life when X-ray and extreme-ultraviolet (EUV) driven escape rates are large and helium and metals are dragged along with the escaping hydrogen. However, as the planet and host star age, diffusive separation of the atmospheric constituents may lead to fractionation and preferential loss of hydrogen\citep{Zahnle1986}.

Hu et al. (2015)\cite{HuEt2015ApJ} first proposed hydrogen-depleted helium-dominated atmospheres on Neptune- and sub-Neptune-size planets to explain the lack of CH$_4$ in GJ~436b’s emission spectrum. Self-consistent calculations of the coupled thermal, mass-loss, and compositional evolution of primordial planetary envelopes\cite{Malsky_2020} have since shown that --- though GJ~436b itself is too large (4.33 $\pm$ 0.18 $R_{\oplus}$\cite{Deming2007}) for photoevaporation to affect its atmospheric composition --- the cumulative effect of preferential loss of hydrogen over billions of years can lead smaller planets $\left(\lesssim3~R_{\oplus}\right)$ to become helium-enhanced.

Here we expand upon the method developed in Malsky \& Rogers (2020)\cite{Malsky_2020}, and show how fractionated mass loss shapes the compositional evolution of the broader planet population of sub-Neptunes. Using the Modules for Experiments in Stellar Astrophysics (MESA v12778)\cite{Paxton2019}, we simulate the evolution of an extensive grid of sub-Neptune primordial envelopes for 10 Gyr. To isolate the effect of mass-loss evolution on planet metallicity, all planets start with a solar composition atmosphere ($X=0.74$, $Y=0.24$, $Z=0.02$).

We predict that planets on the large-radius edge of the radius valley will be enhanced in helium and depleted in hydrogen if the radius valley is primarily produced through photoevaporative mass loss. Figure~\ref{fig:fr-3k-gstar} shows a radius valley forms in our simulations as some planets retain part of their initial hydrogen-helium envelope (at radii $\gtrsim$ 1.6 $R_\oplus$) and some are completely stripped of their atmospheres and become remnant cores (at radii $\lesssim$ 1.9 $R_\oplus$). Billions of years of fractionated atmospheric escape leads to planets that are enhanced in helium and metals relative to their initial conditions, with many planets commonly achieving $Y\geq0.4$. To become helium-enhanced $\left(Y\geq0.4\right)$, planets must undergo extensive mass loss and lose at least 50\% of their initial volatile inventory (by mass), yet necessarily still retain a portion of their initial envelope. Therefore, planets that are helium-enhanced --- and hence have lost most but not all of their primordial envelopes --- fall on the upper edge of the radius valley.

Helium-enhanced planets on the upper edge of the radius valley are a robust outcome of our simulations, and persist for every combination of host star spectral type (G-dwarf, K-dwarf, and M-dwarf) and homopause temperature (ranging from 3,000~K to 10,000~K) that we explored (\S~\ref{sec:homopause_temperature}, \S~\ref{sec:stellar_type}). While increasing the homopause temperature diminishes the level of hydrogen-helium fractionation in the escaping wind, helium-enhanced planets are obtained even at the highest plausible homopause temperature of 10,000~K\cite{MurrayClayEt2009ApJ} after $\gtrsim5$~Gyr. The location of helium enhancement shifts in $M_p-R_p-F_p$ space with host star spectral type, decreasing in radius for planets evolved around lower mass stars. The uncertain parameters in models of photoevaporation (such as the mass-loss efficiency factor) largely shift the location of the radius gap and the helium-enhanced planets in tandem. Thus, benchmarking the predicted helium-enhanced feature in the exoplanet population against the radius gap minimizes the sensitivity to uncertain model parameters.

\begin{figure*}[!htb]
    \includegraphics[width=1.0\linewidth]{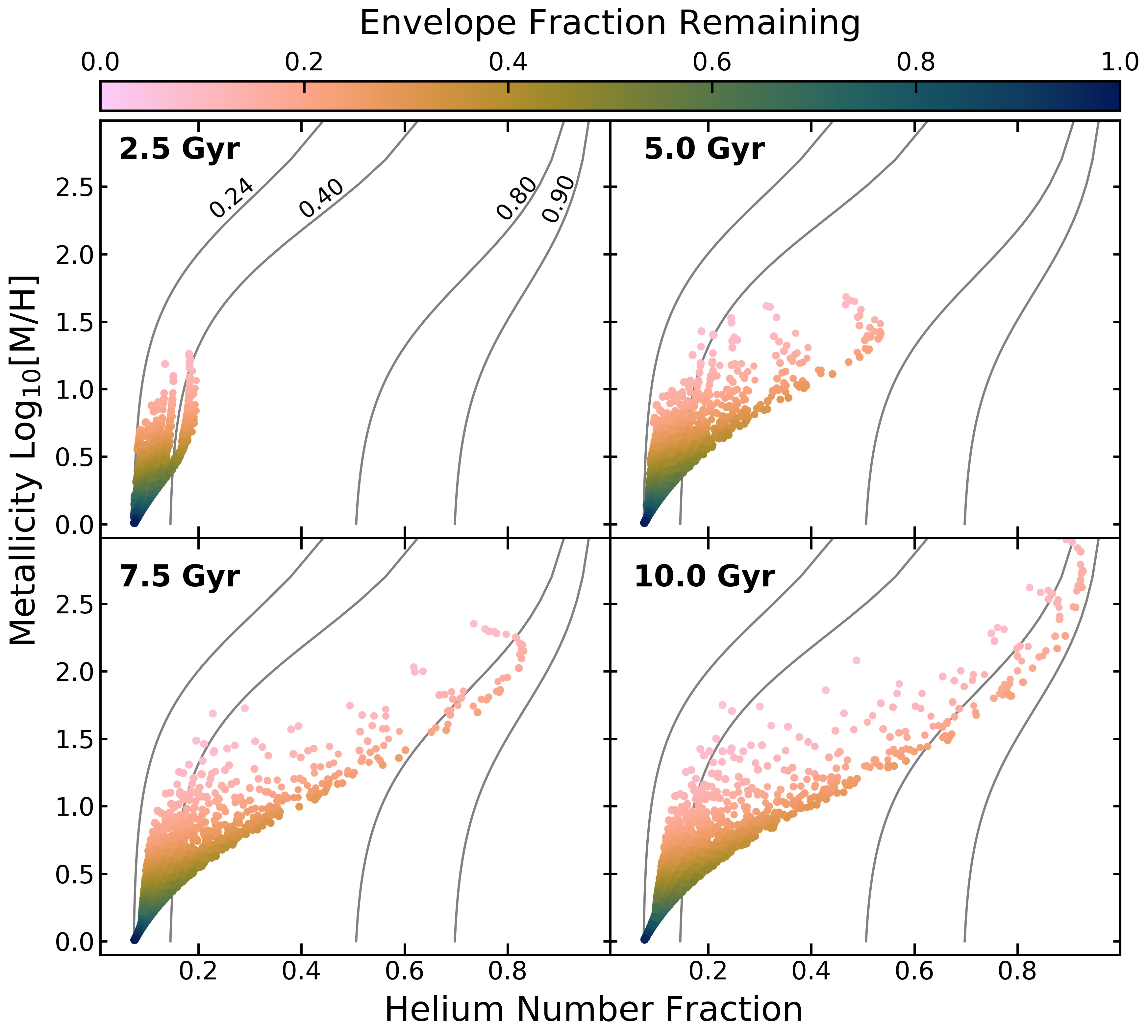}
    \caption{The metallicity and helium number fraction of planets evolved with fractionated mass loss at ages of 2.5, 5.0, 7.5, and 10.0 Gyr. All planets were evolved around a 6,000 K G type host star and began their evolutions with solar compositions  ($X=0.74$,$Y=0.24$, $Z=0.02$). The labeled grey lines show lines of constant helium mass fractions. The spread in envelope compositions at each planet age reflects the differences cumulative mass-loss history across the range of planet masses and orbital separations considered in our grid of simulations. Starting the models with a higher initial envelope metalicity would shift the final envelope compositions to higher metallicities.}
    \label{fig:metals-3k-gstar}
\end{figure*}

The metallicity of a planet's atmosphere carries a signature of its initial formation process (e.g., the timing and size scale of the accretion of solids \cite{FortneyEt2013ApJ}). For planets at the edge of the radius gap, subsequent atmospheric evolution imprints an additional metallicity enhancement. The atmospheres of helium-enhanced planets are dramatically enriched in metals by fractionated atmospheric escape. Atmospheric metallicities, log$_{10}$([Fe/H]), can be enhanced by a factor of 200 over their initial values, though factors of 5 to 30 are more typical (Figure \ref{fig:metals-3k-gstar}).

Fractionated mass loss increases the proportion of helium and metals in sub-Neptune planet atmospheres over timescales of billions of years (Figure \ref{fig:metals-3k-gstar}). Planets must be at least $\sim$ 2 billion years old to accumulate helium envelope mass fractions greater than 0.40. After 2.5 Gyr, most sub-Neptunes still have near-solar atmospheric compositions, but some models with the smallest initial envelopes have $Y\geq0.40$. By 10 Gyr, a wide range of atmospheric compositions are possible, including planets with envelopes that are more than than 80\% helium by mass.

\begin{figure*}[!htb]
    \includegraphics[width=1.0\linewidth]{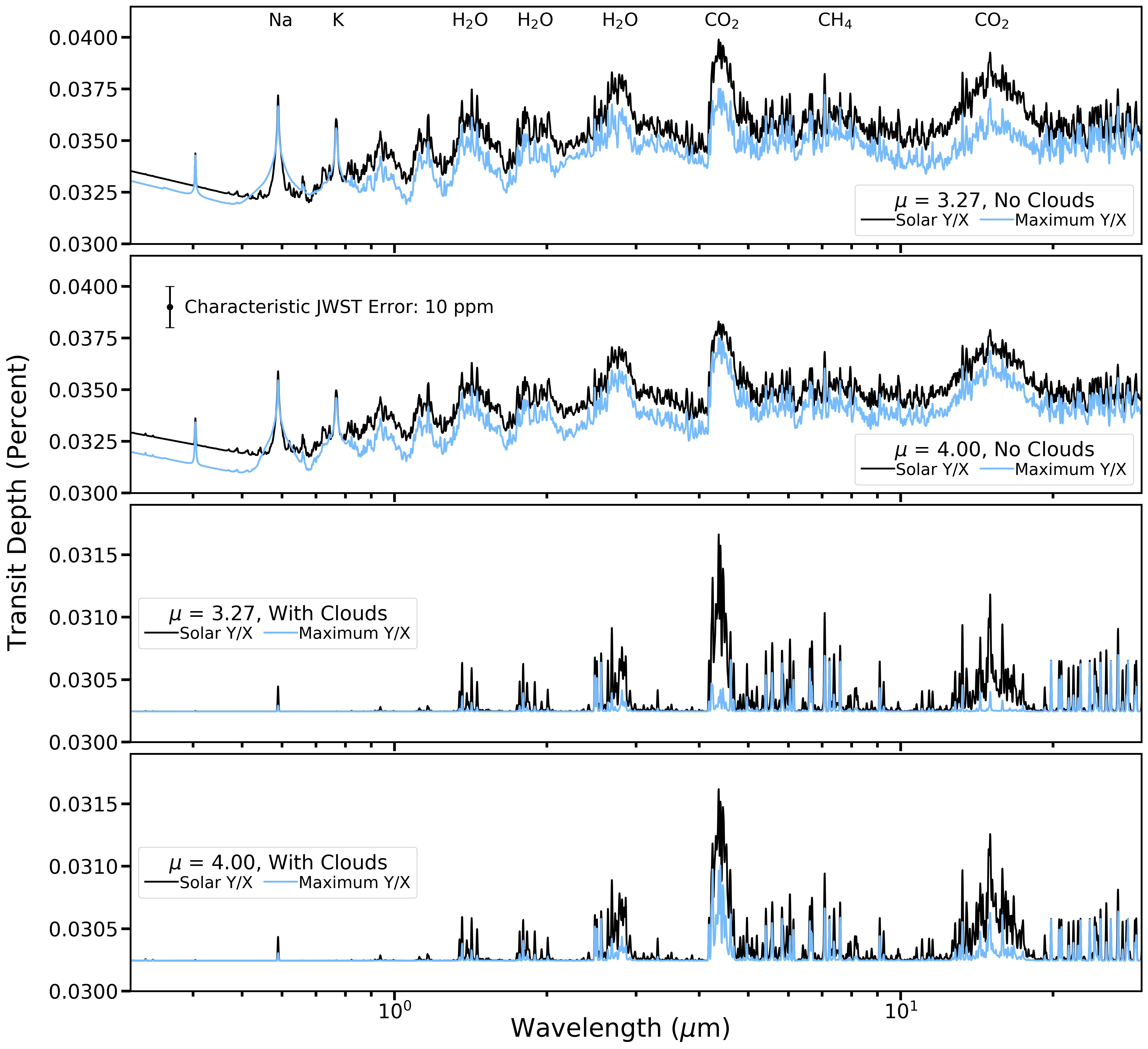}
    \caption{Simulated transmission spectra with varying atmospheric compositions using Exo$\_$Transmit\cite{Kempton2017PASP}. The atmospheric compositions had mean molecular weights of 3.27 (first and third panels) and 4.00 (second and fourth panels). These matched the mean molecular weight of a model with 10x solar metallicity and 100x metallicity respectively, and the maximum Y/X ratio achieved in our model (see Figure \ref{fig:number_fraction_ratios}). Furthermore, the transmission spectra of the top two panels were simulated without clouds. The bottom two panels were simulated with cloud tops at 1.0 Pa, chosen as a representative cloud top pressure for an atmosphere with a solar mean molecular weight\cite{KreidbergEt2014Nature}. The hydrogen, helium, metal number fractions were 0.82, 0.08, 0.10 (bottom black), 0.14, 0.84, 0.02 (bottom blue), 0.86, 0.08, 0.06 (top black), 0.57, 0.42, 0.01 (top blue). When choosing a representative pressure-temperature profile for calculating the transmission spectra we use a planet with a mass of 10.0 $M_\oplus$, an initial envelope fraction of 0.325$\%$, a transit radius of 2.14 $R_\oplus$, a surface gravity of 23.05 m s$^{-2}$, and an orbital separation of 0.09 au. This matches the mean of the distribution of surface gravities, radii, and equilibrium temperatures in our simulations, as shown in Figure \ref{fig:corner}.}
    \label{fig:spectra}
\end{figure*}

Atmospheric helium enhancement has observable consequences for the interpretation of the atmospheric spectra of sub-Neptunes (Figure \ref{fig:spectra}). To date, most atmospheric spectral modeling of primordial envelopes have fixed the ratio of hydrogen and helium to solar abundances; this assumption must now be relaxed. At a constant metal mass fraction, $Z$, increasing the proportion of helium relative to hydrogen (X/Y) increases the atmospheric mean-molecular weight and decreases the atmospheric scale height, which in turn reduces the rate at which the transit depth changes as a function of the extinction cross-section\cite{e-mil09} and is measurable from the shapes of individual spectral features, the relative depths of features from the same molecule, and/or the slope of the Rayleigh signature\cite{Benneke&Seager2012ApJ}.

Comparing atmospheres with identical mean-molecular weights and scale heights (Figure \ref{fig:spectra}), super-solar Y/X cases have lower metallicities (and overall opacities) than solar Y/X cases. Consequently, absorption line cores form deeper in the atmosphere for the helium-enhanced cases, resulting in increased pressure broadening apparent in the width of the sodium and potassium lines. The excess pressure broadening will be more apparent for the single elements lines (especially Na and K) than for the molecular bands, which are blends of many individual vibration-rotation lines.

The relative depth of the molecular Rayleigh scattering signature at short wavelengths, compared to the transit depths from molecular absorption in the near infra-red, measures the mixing ratio of spectrally inactive gasses (e.g., H$_2$, He, and N$_2$) in the atmosphere\cite{Benneke&Seager2012ApJ}. The low Rayleigh cross section of He causes the lower Rayleigh scattering continuum in the maximum Y/X models in Figure \ref{fig:spectra}. Super-solar Y/X may also shift the equilibrium molecular abundances of spectrally active molecules, decreasing the proportion of CH$_4$ relative to CO and CO$_2$ when the number fraction of hydrogen becomes comparable to the number fraction of heavy elements ratio\cite{hu15} (a level of hydrogen depletion only reached by the most extreme outcomes of our simulations).

One complication to using helium enhancement as an observational diagnostic arises if the atmosphere has clouds or haze. For example, some of the biggest spectral differences are expected at shorter wavelengths, where aerosol particles are efficient scatterers. Additional degeneracies between the shape of transmission spectral features and the presence of aerosol layers could significantly complicate the inference of the helium to hydrogen ratio, as shown by the lower panels of Figure \ref{fig:spectra}. However, a detailed study of the degeneracy of clouds and helium enhancement is outside the scope of this work.

Helium has been directly detected in the escaping atmospheres of hot Jupiters (HD189733b, WASP-107b, WASP-69b), warm Neptunes (Hat-P-11b, GJ 3470b), and a young sub-Neptune (TOI 560.01) via absorption in the meta-stable helium 1083 nm line of transmitted starlight\cite{2018ApJ...868L..34M}. Importantly, the fractionation process, whose cumulative effect engenders helium-enhancement in the atmospheres retained by sub-Neptunes, itself causes the escaping winds to be depleted in helium relative to hydrogen. Detecting the time-integrated effects of fractionated escape in the envelopes retained by planets on the upper edge of the radius valley would be complementary to the direct detection of spectral features in the winds currently escaping from sub-Neptunes.

A super-solar abundance ratio of helium relative to hydrogen is an observable signature of a planetary envelope of primordial origin that has been sculpted by hydrogen loss (via atmospheric escape and/or H$_2$-magma interactions). The ratio of helium to hydrogen in planet-forming disks was set by primordial Big Bang nucleosynthesis and has not been significantly modified since\cite{CocEt2015PRD}. As a non-reactive noble gas, helium is not incorporated into minerals or ices and thus cannot be accreted by a planet in the form of rocky or icy solids\cite{ElkinsTanton&Seager2008aApJ}. Due to the low relative cosmic abundances of unstable radioactive nuclides, the amount of helium produced by alpha decays is negligible compared to the helium-enhanced envelopes in our simulations (wherein helium accounts for $\sim 0.02\%$ of the planet mass).  Thus, outgassing and delivery of volatiles by icy pebbles or planetesimals will only dilute the helium-to-hydrogen ratio in planetary atmospheres.

Not all planets on the upper edge of the radius gap will necessarily be helium-enhanced. Planets that are less than a few billion years old will not have time to accumulate the effects of preferential hydrogen loss and water worlds may also be possibilities in this parameter space\cite{ZengEt2019PNAS}. However, atmospheric helium enhancement presents an important avenue for testing the origins of the radius valley. An observational detection of helium-enhanced planets on the upper edge of the radius valley would break the degeneracy between sub-Neptune planet compositional scenarios, and provide insights into the formation and evolution of this enigmatic and abundant planet population. 
 
\section*{Code availability}
MESA is publicly available (http://mesa.sourceforge.net/). Exo$\_$Transmit is also publicly available (https://github.com/elizakempton/Exo$\_$Transmit).

\appendix

\section{Appendix Methods}\label{sec:more-methods}
To model the coupled thermal, mass-loss, and compositional evolution of primordial envelopes surrounding sub-Neptune mass planets, we use the Modules for Experiments in Stellar Astrophysics (MESA v12778)\cite{Paxton2011, Paxton2013, Paxton2015, Paxton2018, Paxton2019}. We follow the modeling approach from Malsky \& Rogers (2020)\cite{Malsky_2020} with several additions. First, we now self-consistently model the hydrogen ionization fraction used to calculate the rate of momentum exchange between hydrogen and helium. Second, we have updated our atmospheric boundary conditions within MESA.

For each simulated planet, we create an initial MESA planet model with the desired combination of initial total planet mass, core mass, atmospheric composition, and entropy. All models begin with a solar composition primordial envelope surrounding a rocky core with solar proportions of silicates and iron. The $M_p-R_p$ relation of Earth-composition rocky cores\cite{2011ApJ...738...59R} sets the inner boundary condition of the MESA model of the hydrogen-helium envelope. To set the core luminosity, we assume\cite{GuillotEt1995ApJ} that the rocky core has a heat capacity of $c_v$=1.0 J K$^{-1}$ g$^{-1}$, and include the contribution from the decay of radionuclides, following Chen \& Rogers (2016)\cite{Chen&Rogers2016ApJ}.

\subsection{Atmospheric Boundary Conditions}
To set the boundary conditions and atmospheric profile, we model the atmosphere up to a optical depth (from the planet's local thermal irradiation) of $\tau$=2/3, and implement a grey Eddington T($\tau$) relation with the \verb|atm_T_tau_relation| option in MESA. We define planetary transit radius (R$_p$) to be the location where the pressure is equal to 1.0 mbar. This roughly corresponds to the radii observed by transit surveys\cite{2009ApJ...702.1413M, MillerEt2009ApJ}. To extrapolate to pressures below the outermost zone in MESA (at approximately 80 millibar) we assume an isothermal temperature profile and a constant value for the mean molecular mass of the atmosphere. As in Malsky \& Rogers (2020)\cite{Malsky_2020}, we use gaseous mean opacities from Freedman et al. (2014)\cite{Freedman2014} and model irradiation from the host star by specifying both the incident stellar flux and the column depth that the flux penetrates down to in the planet's atmosphere.

We standardize the initial thermal profile of the planet at the beginning of evolution to a ``hot start''. At the start of the evolution stage the planet envelope cools and gravitationally contracts on a Kelvin-Helmholtz timescale. Over 6.0 Myr the irradiation from the planet's host star is increased from 0 to the full specified irradiation. At 6.0 Myr, the planet has been brought to the correct starting state and begins fractionated mass loss.

We define the homopause as the location where the hydrogen-helium binary diffusion coefficient is equal to the eddy diffusion coefficient. Below the homopause radius, turbulence and convection homogenize the planet atmosphere. Above the homopause radius, fractionation of hydrogen and helium can lead to differences in atmospheric abundances. Generally, the homopause radius of the planet is approximately 10\% larger than the transit radius of a planet. Throughout this work we adopt a value of K$_{zz}$ = 10$^9$ cm$^2$ s$^{-1}$ for the eddy diffusion coefficient. Increasing (decreasing) the eddy diffusion coefficient by a factor of 10 results in approximately a 5\% larger (smaller) homopause radius\cite{Malsky_2020}.

\subsection{Photoevaporation}
During evolution, planets lose mass due to photoevaporation driven by EUV radiation\cite{MurrayClayEt2009ApJ, 2012MNRAS.425.2931O}. Ionizing stellar EUV radiation heats the outer layers of the planet envelope (via thermalization of electrons ionized from hydrogen atoms) and drives a hydrodynamic wind from the planet. The EUV flux from the star, which drives the mass loss, decreases exponentially in time as the star evolves. We parameterize the star's EUV luminosity following the equations in Sanz-Forcada et al. (2011)\cite{2011A&A...532A...6S} and model fractionated mass loss from photoevaporation following an approach adapted from Hu et al. (2015)\cite{HuEt2015ApJ}. We use $\Phi$ to denote the total mass loss rate from the planet (mass per time), and $\phi$ to denote the number fluxes of particles escaping from the planet (particles per area per time).

At low EUV fluxes, the overall mass loss is approximated as energy-limited, wherein a fixed fraction of the EUV luminosity impinging on the planet contributes to unbinding mass from the gravitational potential well of the planet. The energy-limited mass-loss rate is
\begin{equation}\label{HeliumEq}
\Phi_{\rm EL} = \frac{L_{\rm EUV}\eta a^2 R_{h}^3}{4Kd^2GM_{p}},
\end{equation}
\noindent where $L_{\rm{EUV}}$ is the EUV luminosity, $M_p$ is the mass of the planet, R$_{h}$ is the homopause radius, $K$ is the Roche potential reduction factor\cite{Roche}, $\eta$ is the heating efficiency, $d$ is the orbital separation, and $a$ is the ratio between the EUV absorbing radius and the homopause radius. EUV photons are deposited at a radius corresponding to approximately $\tau_{\rm{EUV}}$=1, which places the EUV absorbing radius within 10$\%$ of the planet's homopause radius\cite{2002ASPC..269..133R,f0fa6183aaa547d4b6c235787bb17e4f, 2009ApJ...693...23M, 2014RSPTA.37230089K, HuEt2015ApJ}. When calculating the energy-limited mass loss rate, we adopt $a=1$ following Hu et al. (2015) to subsume the uncertainty in the ratio between the EUV absorbing radius and the homopause radius into other parameters (namely $\eta$) in the energy-limited escape formulation.

While the energy-limited escape rate is a good approximation when the escaping wind is subsonic, simulations show that it breaks down when the flow is transonic\cite{Tucker2012, Erwin2013, JohnsonEt2013ApJ, Volkov2013}. At large EUV heating rates (Q$_{net}$ $\gtrsim$ 5$\times$10$^{13}$ - 5$\times$10$^{14}$ ergs s$^{-1}$ in our simulations) the majority of the incident radiation is converted into translational and thermal energy in the atmosphere and the mass loss becomes less efficient. For planets receiving EUV fluxes above the critical minimum heating rate to drive a transonic flow, the mass loss rate saturates and no longer increases with energy input.  In this transonic escape regime, we modify the energy-limited escape rate with the efficiency reduction factor, $f_r$, from  Johnson et al. (2013)\cite{JohnsonEt2013ApJ}, $\Phi=f_r\Phi_{\rm EL}$.

\subsection{Fractionation}
At radii above the homopause, atmospheric constituents separate out by their molecular weight, with the lighter species extending out to higher altitudes due to their larger atmospheric scale heights. The diffusive separation of atmospheric constituents leads heavier species (helium and metals) to be preferentially retained as the hydrogen is lost. It is thus convenient to separate the total mass loss rate, $\Phi$, into the separate contributions from hydrogen and helium escape
\begin{equation}
    \Phi = \Phi_{\rm H}+\Phi_{\rm He} = 4\pi R_{h}^2\left(\phi_{\rm H}m_{\rm H}+\phi_{\rm He}m_{\rm He}\right),
    \label{eq:Phitot}
\end{equation}
where $\Phi_{\rm H}$ and $\Phi_{\rm He}$ are the mass loss rates of hydrogen and helium, and $\phi_{\rm H}$ and $\phi_{\rm He}$ are the fluxes of hydrogen and helium particle escaping per unit time and per unit area.

The diffusion of helium relative to the escaping hydrogen is characterized by an effective binary diffusion coefficient (b$^\prime$) as
\begin{eqnarray}
\frac{k \rm{T_H}}{b'} = (1 - x)\frac{k \rm{T_H}}{b} + x\frac{m_{H} \nu }{n_{\rm {He}}}
\label{eq:bprime}
\end{eqnarray}
\noindent where $b = 1.04 \times 10^{18} \rm{T^{0.732}} \rm{cm^{-1} s^{-1}}$ is the binary diffusion coefficient between neutral hydrogen and helium\cite{Mason1970}, $\nu$ is the ion-neutral momentum transfer collision frequency,\cite{schunk1980ionospheres} and $x$ is the ionization fraction of hydrogen. The first term on the right hand side of Eq~\ref{eq:bprime} reflects the coupling between H and He, while the second term on the right hand side represents the coupling between H$^{+}$ and He. We improve upon Malsky \& Rogers (2020)\cite{Malsky_2020}, which assumed a constant hydrogen ionization fraction of 0.1\cite{HuEt2015ApJ,Malsky_2020}, by calculating the ionization fraction at the homopause radius at each timestep. This allows us to model the fractionation between hydrogen and helium for varying homopause temperatures and pressures.

The fractionated escape fluxes of hydrogen and helium particles are approximated as

\begin{equation}
    \frac{\phi_{\rm He}}{X_{\rm He}}=\frac{\phi_{\rm H}}{X_{\rm H}}-\frac{G M_p (m_{\rm He} - m_{\rm H}) b'}{R_h^2 k T_{H}},
    \label{eq:fract}
\end{equation}

\noindent where $X_{\rm H}$ and $X_{\rm He}$ are the mixing ratios of hydrogen and helium at the homopause, $k$ is the Boltzmann constant, m$_{\rm H}$ is the mass of a hydrogen atom, and m$_{\rm He}$ is the mass of a helium atom. The second term on the right hand side of equation~\ref{eq:fract} is denoted by Hu et al. (2015) as the diffusion-limited escape rate $\phi_{\rm DL}$,

\begin{equation}
\phi_{\rm DL} = \frac{G M_p (m_{\rm He} - m_{\rm H}) b'}{R_h^2 k T_{H}}. 
\label{eq:phiDL}
\end{equation}

We note that this definition differs slightly from the diffusion-limiting flux of hydrogen escaping through a stationary background atmosphere defined by Hunten (1973)\cite{Hunten1973JAtS}, which is related to the expression in Equation~\ref{eq:phiDL} by $X_{\rm H}\phi_{\rm DL}$.

Equations~\ref{eq:Phitot} and \ref{eq:fract} together reveal that the extent of the fractionation is divided into two regimes determined by the mass loss rate. When the mass loss rate is large compared to the diffusion-limited escape rate $\left(\phi_{\rm H}/X_{\rm H}\gg\phi_{\rm DL}\right)$, hydrogen and helium are lost in approximately equal proportion to their mixing ratios at the homopause radius. During this rapid evaporation stage, hydrogen and helium are strongly coupled and relatively little helium enhancement occurs. As the escape rate decreases and approaches the diffusion limited mass escape rate of hydrogen, the escaping wind from the planet becomes more and more enriched in hydrogen relative to helium. Once the mass loss rate decreases below the critical diffusion limited mass loss rate, only hydrogen escapes. After each time step in the evolution of the MESA planet model, we update composition of the remaining envelope that was retained by the planet to reflect the differing amounts of hydrogen and helium that were lost. This is accomplished as part of the \verb|extras_finish_step| routine, as described in Malsky \& Rogers (2020)\cite{Malsky_2020}.

\subsection{Grid Sub-Neptune Evolution Models}
The grid of planets modeled has 17 masses from 4.0 to 20.0 M$_\oplus$, 25 initial envelope mass fraction from 0.001 to 0.01, and 30 orbital separations from 0.01 to 0.3 au. Additionally, for each of these parameterizations we model planets orbiting G stars with T$_{eff}$ = 6,000 K, M$_\star$ = 1.0 M$_\odot$, and R$_\star$ = 1.0 R$_\odot$, K stars with T$_{eff}$ = 4,780, M$_\star$ = 0.75 M$_\odot$, and R$_\star$ = 0.73 R$_\odot$, and M stars with T$_{eff}$ = 3,600 K, M$_\star$ = 0.2 M$_\odot$, and R$_\star$ = 0.30 R$_\odot$. For each set of models we simulated homopause temperatures of 3,000 K and 10,000 K for a total of 76,500 planet models.

\subsection{Chemistry of Helium-Enhanced Atmospheres}
\begin{figure*}[!htb]
    \includegraphics[width=1.0\linewidth]{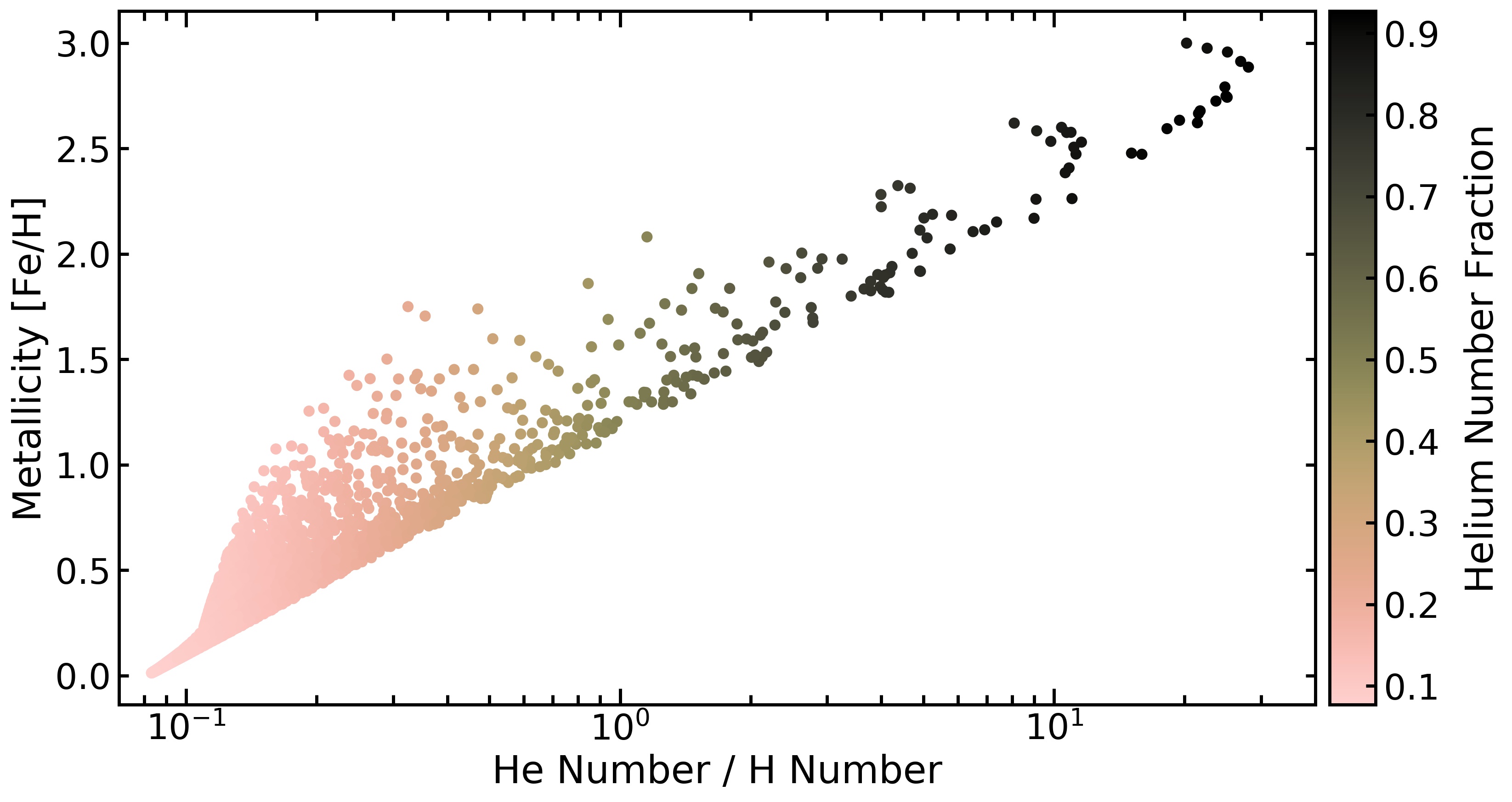}
    \caption{The number fraction ratios of all planets evolved for a parameterization identical to that used for Figure \ref{fig:mr-3k-gstar}. The colored dots correspond to the compositions reached at 10 Gyr.}
    \label{fig:number_fraction_ratios}
\end{figure*}

\begin{figure*}[!htb]
    \includegraphics[width=1.0\linewidth]{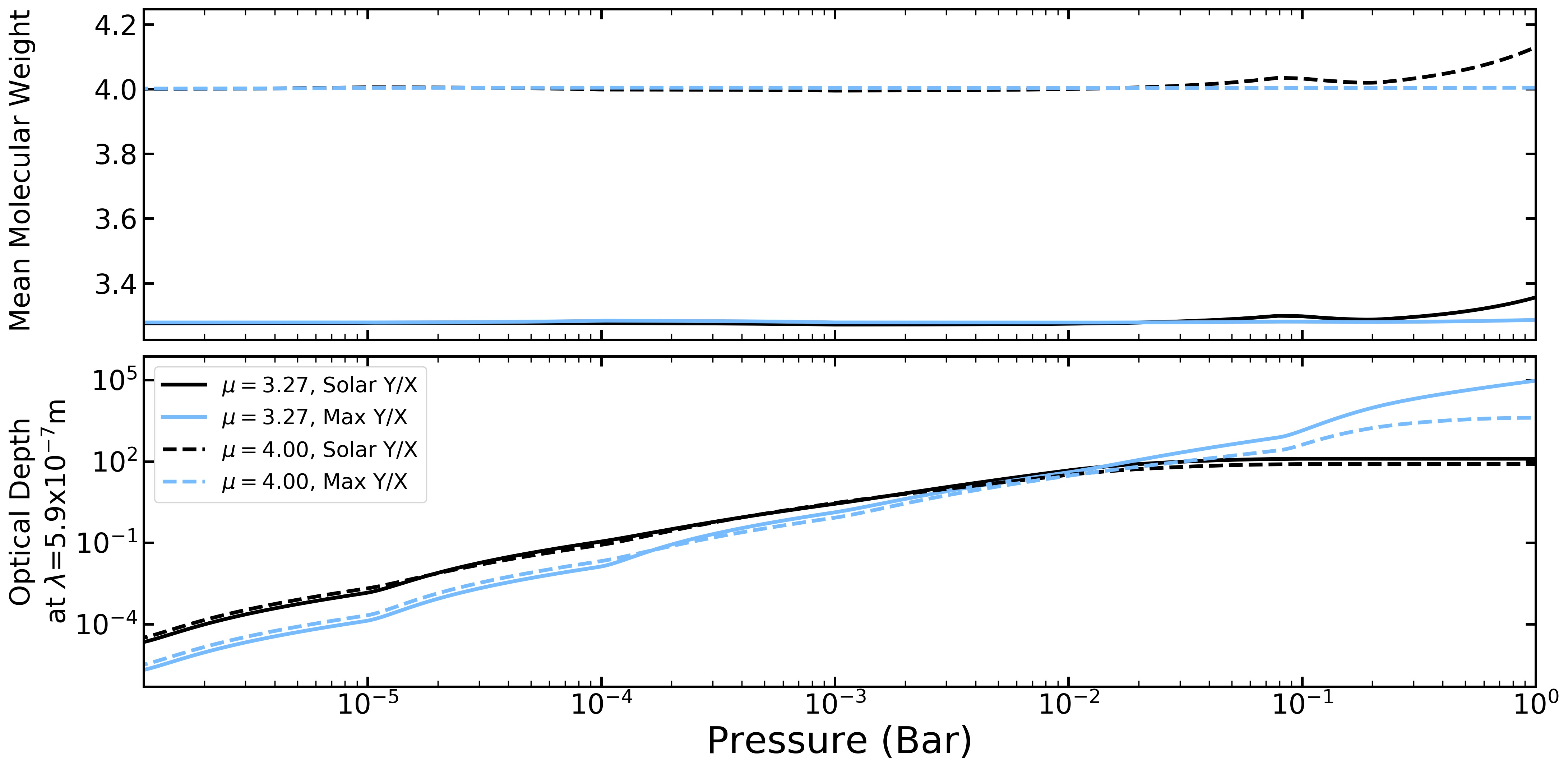}
    \caption{The mean molecular weights as a function of pressure (top panel) and the optical depth (at $\lambda=5.9\times10^{-7}$~m, near the rest wavelength of the Na D-lines) as a function of pressure (bottom panel) for the solar Y/X composition atmospheres and the maximum Y/X composition atmospheres.}
    \label{fig:mmw}
\end{figure*}

In order to understand how helium enhancement manifests in observations, we simulate transmission spectra for a number of atmospheric compositions. Figure \ref{fig:number_fraction_ratios} shows the spread in compositions after 10 Gyr of mass loss. We selected compositions with highest helium/hydrogen enhancement at metallicities of 10x solar and 100x solar. Next, we took the relative abundances of atmospheric constituents from Lodders (2003)\cite{Lodders2003} and scaled them to the helium and metal enhancements of our two selected models. Then, we found atmospheric compositions with solar helium to hydrogen ratios that matched the mean molecular weight of the 10x solar and 100x solar metallicity helium enhanced models, as shown in Figure \ref{fig:mmw}.

\begin{figure*}[!htb]
    \includegraphics[width=1.0\linewidth]{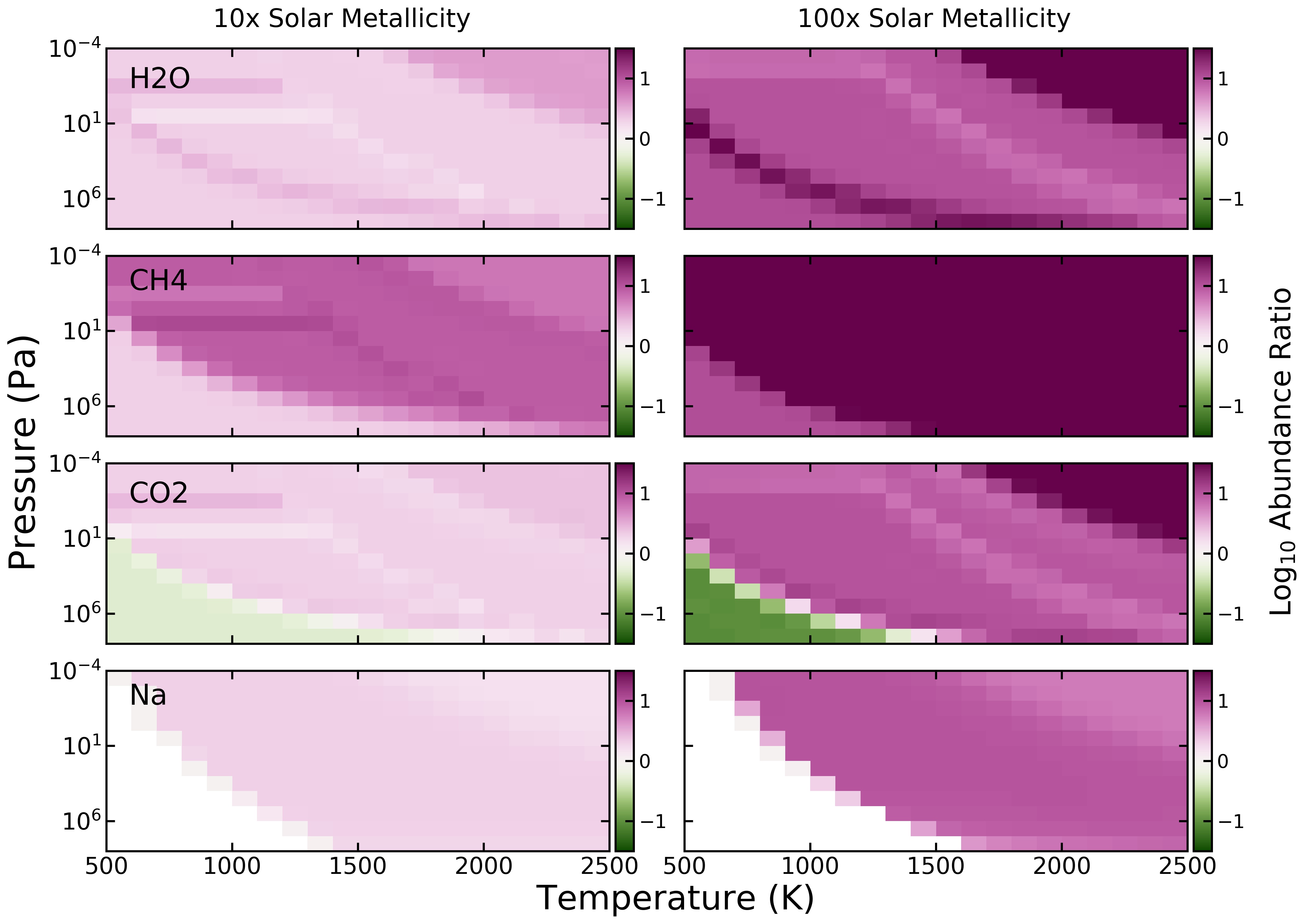}
    \caption{The abundance ratios of the solar Y/X composition EOS tables divided by the maximum Y/X composition EOS tables at 10x solar metallicity and 100x solar metallicity for H$_2$O, CH$_4$, CO$_2$, and Na. These tables show the EOS abundances at different X/Y ratios but constant metallicities.}
    \label{fig:abundances}
\end{figure*}

\begin{figure*}[!htb]
    \includegraphics[width=1.0\linewidth]{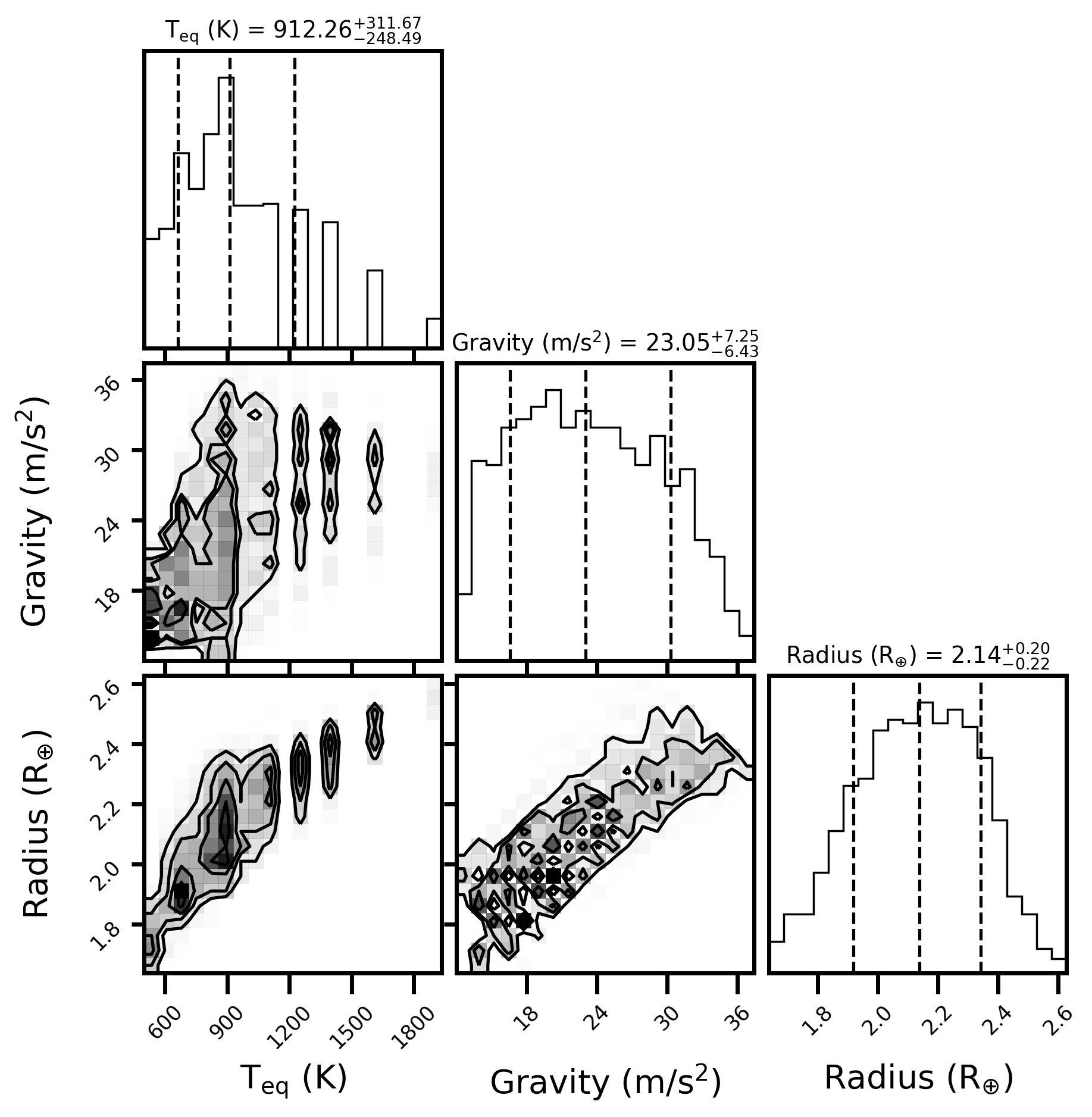}
    \caption{The distribution of surface gravities, radii, and equilibrium temperatures for all planets with Y $\geq$ 0.4 in our simulations after 10 Gyr. All model parameters are identical to those in Figure \ref{fig:mr-3k-gstar}.}
    \label{fig:corner}
\end{figure*}

We calculate abundances in thermochemical equilibrium for the most important atmospheric absorbers over a grid of temperature and pressure (i.e. equation of state (EOS) tables in Exo$\_$Transmit format for representative X / Y / Z compositions in our model grid), using the methods outlined in Mbarek \& Kempton (2016)\cite{Mbarek2016}. The abundances of key species are shown in Figure \ref{fig:abundances}. Importantly, these new EOS tables highlight that the helium enhanced atmospheres have much lower metallicities for constant mean molecular weights. We benchmarked our solar helium to hydrogen ratio tables against the ones included in Exo$\_$Transmit and found perfect agreement. Finally, we choose a representative temperature-pressure profile for the distribution of surface gravities, radii, and equilibrium temperatures of helium enhanced planets found in our simulations (Figure \ref{fig:corner}) and extrapolate an isothermal upper atmosphere extending from the transit radius to 0.1 mbar to calculate transmission spectra.

\section{Supplemental Results}
\subsection{Candidates for Helium Enhancement}
\begin{figure*}
    \includegraphics[width=1.0\linewidth]{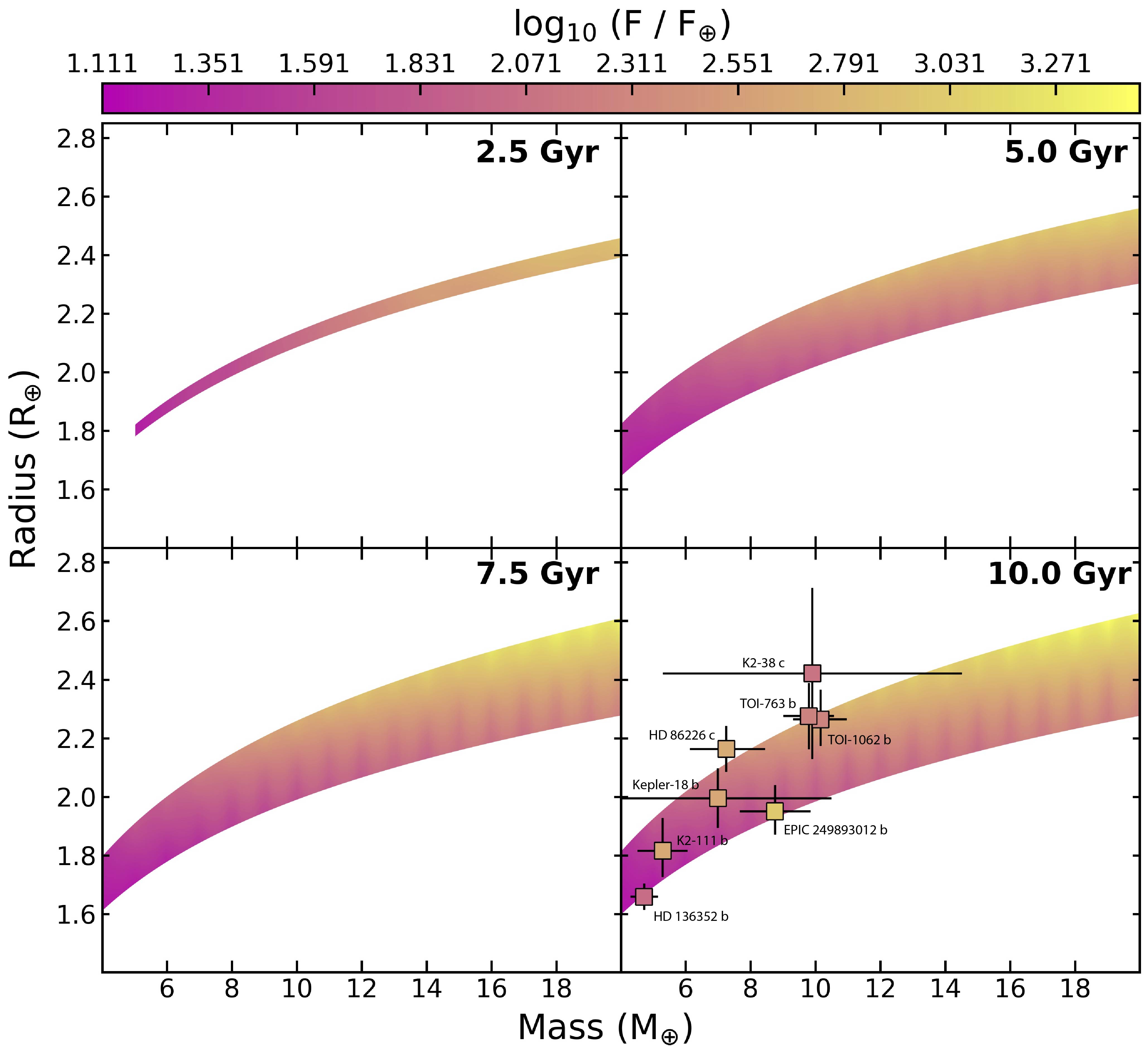}
    \caption{The mass-radius-flux parameter space for helium enhanced planets orbiting G stars. The panels show the parameters space at ages of 2.5, 5.0, 7.5, and 10.0 Gyr. The color shading corresponds to the minimum instellation required for helium enhancement in our simulations. The square points correspond to observed exoplanets and their error-bars. Helium enhancement may be possible for planets with $M_p$ $\leq$ 4 $M_\oplus$. However, below 4 $M_\oplus$ many planets exceeded the $\rho$ - T boundary limits in the EOS module of MESA and could not be simulated. Broader simulation of different parameters of our model (stellar host type, homopause temperature, escape efficiency, etc.) may mean that other planets overlap with the helium enhancement zone depicted here.}
    \label{fig:mr-3k-gstar}
\end{figure*}

Figure \ref{fig:mr-3k-gstar} shows the 3-dimensional volume of planetary mass-radius-incident flux ($M_p-R_p-F_p$) parameter space in which helium enhancement is found. Helium-enhanced planets reside in a narrow arc of $M_p-R_p-F_p$ parameter space with radii between 1.6 and 2.5 R$_\oplus$, incident flux rates between 10 F$_\oplus$ and 800 F$_\oplus$, and masses from 4.0 to 20.0 M$_\oplus$. As planets age and lose hydrogen preferentially the parameter space for helium enhancement expands.

To determine the mass-radius-flux parameter space for helium enhanced planets (as shown in Figure \ref{fig:mr-3k-gstar}) we found the minimum flux necessary for helium enhancement at each mass and radius. Because of differences in initial envelope mass fractions, there may be multiple helium enhanced model evolution tracks that lead to the same planet mass and radius at a given age. First, for each host star type and homopause temperature, we filter the population of sub-Neptunes to include only planets that have Y $\geq$ 0.4. Next, we interpolated over the filtered models using a radial basis function\cite{2020SciPy-NMeth} to find the flux at each point in the mass-radius parameter space and fit the upper and lower $M_p-R_p$ relations with logarithmic functions as they best matched the bounds of our modeled planets. Planets were included as helium enhancement candidates if they had fluxes between 0.1 and 10x the flux value of our $M_p-R_p$ interpolation. The flux value for each planet was calculated using parameters from the NASA Exoplanet Archive\cite{ExoDatabase}.

To prioritize helium enhanced exoplanet candidates based on observability, we calculate a transmission spectroscopy metric\cite{kempton18} (TSM) score for each planet, shown in Table \ref{table:candidates}. This score is a measure of the quality of a candidate for atmospheric characterization, with higher scores meaning that a planet is more readily accessible:

\begin{equation}
\rm{TSM} = (\rm {Scale \ factor}) \times \frac{R_p^3 T_{eq}}{M_p R_\star^2}\times 10^{-m_J/5}
\end{equation}

\noindent where R$_\star$ is stellar radius, $T_{eq}$ is the planet's equilibrium temperature assuming zero albedo, and m$_J$ is J band apparent magnitude of the host star. We adopt a scale factor of 1.26 for planets with radii between 1.5 R$_\oplus$ and 2.75 R$_\oplus$\cite{kempton18}.

Among the exoplanets discovered orbiting G, K, or M stars, there are a number of candidates for helium-enhanced atmospheres. Table \ref{table:candidates} shows the relevant properties for each planet in the $M_p-R_p-F_p$ parameter space for which we predict helium enhancement after 10 Gyr of fractionated mass loss. The measured properties of these planets overlap (within their 1-sigma measurements uncertainties) with the $M_p-R_p-F_p$ parameter space in which we find helium-enhanced planets. There are a number of candidates around G and K star planets. However, as yet we found no candidates for helium enhancement around M stars.

\begin{deluxetable*}{llllll}\label{table:candidates}
\tablehead{\colhead{Name} & \colhead{Mass (M$_\oplus$)} & \colhead{Radius (R$_\oplus$)} & \colhead{Flux (F$_\oplus$)} & \colhead{J Band Magnitude} & \colhead{TSM}}
\startdata
\hline
\hline
\multicolumn{6}{c}{\textit{G Star Planets}}\\
\hline
\hline
EPIC 249893012 b     &  8.75 &         1.95 &       1032.49 &        10.22 &        26.74 \\
HD 136352 b          &  4.72 &         1.66 &       111.65  &        4.31  &      601.73 \\
HD 86226 c           &  7.25 &         2.16 &       486.81  &        6.84  &      618.47 \\
K2-111 b             &  5.29 &         1.82 &       479.92  &        9.77  &       56.06 \\
K2-38 c              &  9.90 &         2.42 &       128.32  &        9.91  &      131.29 \\
Kepler-18 b          &  6.99 &         2.00 &       451.56  &       12.19  &       34.30 \\
TOI-1062 b           & 10.15 &         2.26 &       188.67  &        8.78  &      447.62 \\
TOI-763 b            &  9.79 &         2.28 &       178.10  &        8.86  &      350.75 \\
\hline
\hline
\multicolumn{6}{c}{\textit{K Star Planets}}\\
\hline
\hline
TOI-1235 b            & 5.90 &          1.69  &        60.13  &         8.71     &    360.17 \\
TOI-1749 c           & 14.00 &          2.12   &       34.88     &     11.07   &      298.06 \\
TOI-178 c            &  4.77  &         1.67   &       96.08     &      9.37    &     249.03 \\
\hline
\hline
\multicolumn{6}{c}{\textit{M Star Planets}}\\
\multicolumn{6}{c}{\textit{\rm{None}}}\\
\hline
\hline
\enddata
\caption{All planets in the $M_p-R_p-F_p$ parameter space for which we predict helium enhancement after 10 Gyr of fractionated mass loss. Masses, radii, and orbital separations are all taken from the NASA Exoplanet Archive\cite{ExoDatabase}. The data were retrieved on February 14th, 2022. Planets with radii larger than 1.5 R$_\oplus$ and TSM scores above 90 are high quality candidates for atmospheric characterization\cite{kempton18}. We use the SAG 13 definitions for the effective temperature cutoffs for G, K, and M stars.}
\end{deluxetable*}

\subsection{Homopause Temperature}\label{sec:homopause_temperature}
As the temperature at the homopause increases, the effects of fractionation decrease. First, the coupling of neutral hydrogen and helium increases with increasing temperature\cite{Mason1970, schunk1980ionospheres}. Second, hydrogen ionization increases with increasing temperature, and ionized hydrogen is more strongly coupled with helium than neutral hydrogen.

In order to quantify how helium enhancement changes with homopause temperature (T$_{\rm{H}}$), we simulate planet evolution with a lower estimate (3,000 K), and an upper estimate (10,000 K). Homopause temperatures above 10,000 K are unphysical as Lyman-$\alpha$ cooling thermostats the upper atmosphere temperatures\cite{MurrayClayEt2009ApJ}. Previous work simulating the thermospheres of sub-Neptunes have used 3,000 K as a lower bound\cite{salz16}, and cooler values would only further increase the level of helium enhancement in sub-Neptunes.

\begin{figure*}[!htb]
    \includegraphics[width=1.0\linewidth]{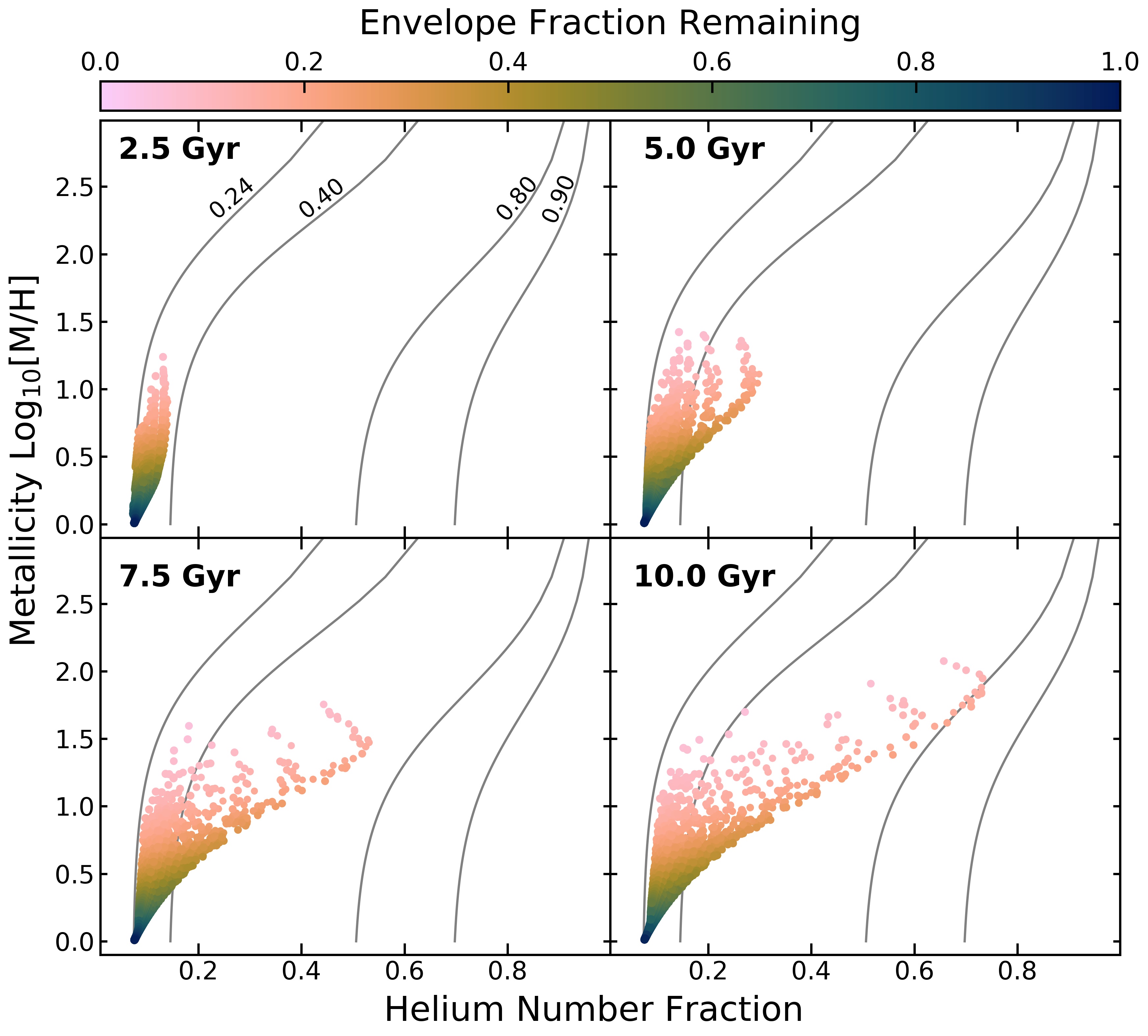}
    \caption{The metallicity and helium number fraction of planets with a homopause temperature of 10,000 K and evolved with fractionated mass loss at ages of 2.5, 5.0, 7.5, and 10.0 Gyr around a G type star. All model parameters other than homopause temperatures are identical to Figure \ref{fig:metals-3k-gstar}.}
    \label{fig:metals-10k-gstar}
\end{figure*}

\begin{figure*}[!htb]
    \includegraphics[width=1.0\linewidth]{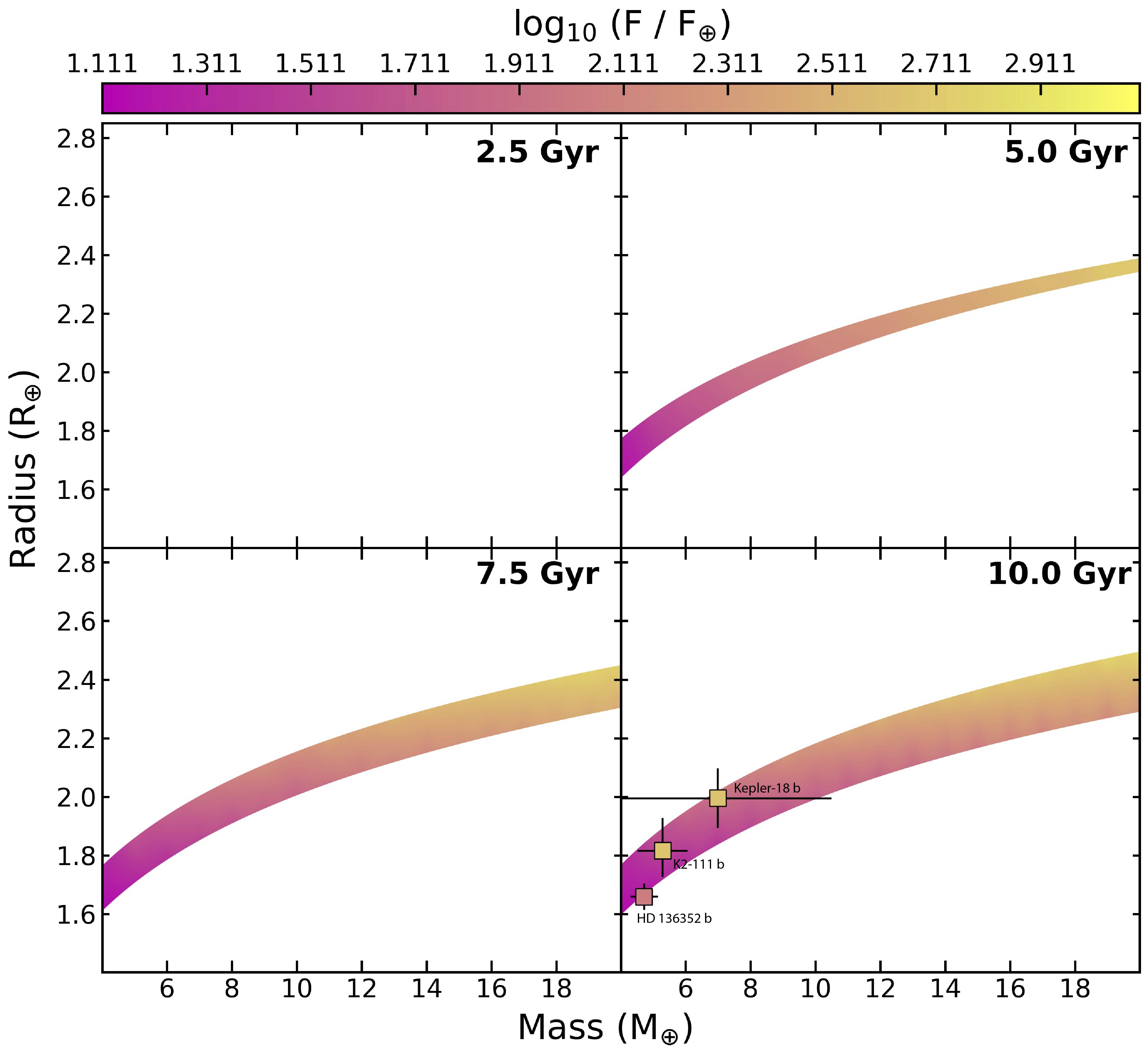}
    \caption{
    The mass-radius relationship for helium enhanced planets with homopause temperatures of 10,000 K at ages of 2.5, 5.0, 7.5, and 10.0 Gyr around a G type star. All model parameters other than homopause temperatures are identical to Figure \ref{fig:mr-3k-gstar}.}
    \label{fig:mr-10k-gstar}
\end{figure*}

Figure \ref{fig:metals-10k-gstar} shows a population of simulated planets evolved with homopause temperatures of 10,000 K. Compared to Figure \ref{fig:metals-3k-gstar}, which shows planets simulated with homopause temperatures of 3,000 K, these planets have less extreme helium and metal enhancement, and became helium enhanced at older ages (generally after 5 Gyr). Nonetheless, many of these planets still attain atmospheric helium mass fractions greater than 0.40 and even as extreme as 0.80. The robustness of helium enhancement for higher homopause temperatures is also paralleled in Figures \ref{fig:fr-10k-gstar} and \ref{fig:mr-10k-gstar}, which show the flux-radius and mass-radius relationship for helium enhanced planets with homopause temperatures of 10,000 K.

\subsection{Stellar Type}\label{sec:stellar_type}
Helium enhancement is a prominent feature of populations of sub-Neptune mass planets that evolve around G, K, and M type stars. Host star spectral type affects the mass-loss history of planets (at specified initial planet mass, envelope mass fraction, and irradiation flux) in two ways. Lower mass stars have a higher ratio of EUV luminosity to total bolometric luminosity (with L$_{\rm{EUV}}$ /  L$_{\rm{BOL}}$ equal to $\sim$ 4, 9, and 70 at 5 Gyr for our simulated G, K, and M stars respectively). Stellar tidal forces are a second factor contributing to differences in planet evolution tracks with host star spectral type. For the same instellation, $F_p$, planets orbiting lower mass, less luminous stars have closer orbital separations $d$ and smaller Roche lobe radii. The closer proximity of the Roche lobe boundary to the planet further enhances the mass loss rates for planets orbiting K or M stars compared to those orbiting G stars. Planets with lower mass host stars have smaller Roche potential reduction factors $K$\cite{Roche}, which in turn increases the mass loss rate $\Phi_{\rm EL}$ in Equation~\ref{HeliumEq}.

\begin{figure*}[!htb]
    \includegraphics[width=1.0\linewidth]{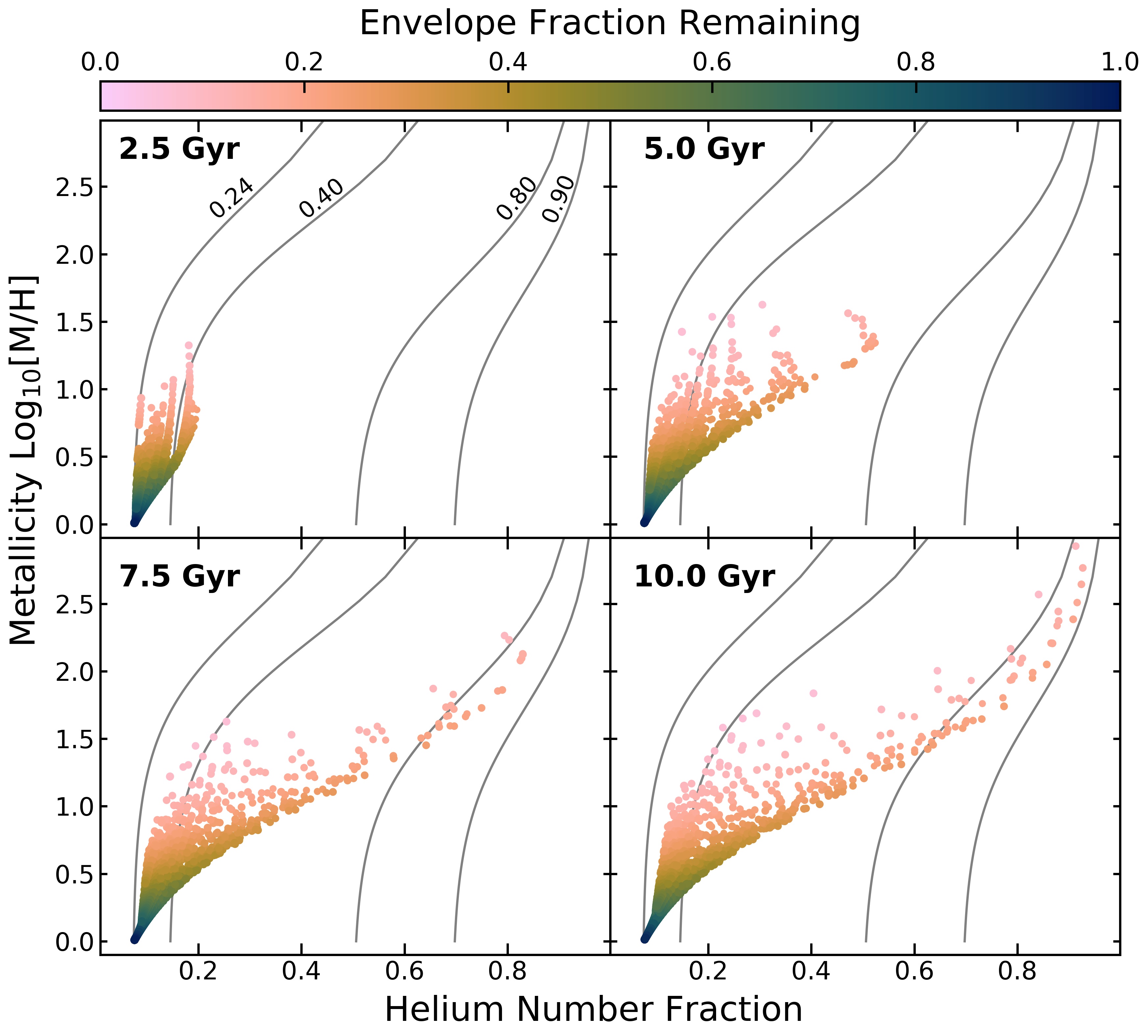}
    \caption{The metallicity and helium number fraction of planets with a homopause temperature of 3,000 K and evolved with fractionated mass loss at ages of 2.5, 5.0, 7.5, and 10.0 Gyr around a K type star. All model parameters other than host star types are identical to Figure \ref{fig:metals-3k-gstar}.}
    \label{fig:metals-3k-kstar}
\end{figure*}

\begin{figure*}[!htb]
    \includegraphics[width=1.0\linewidth]{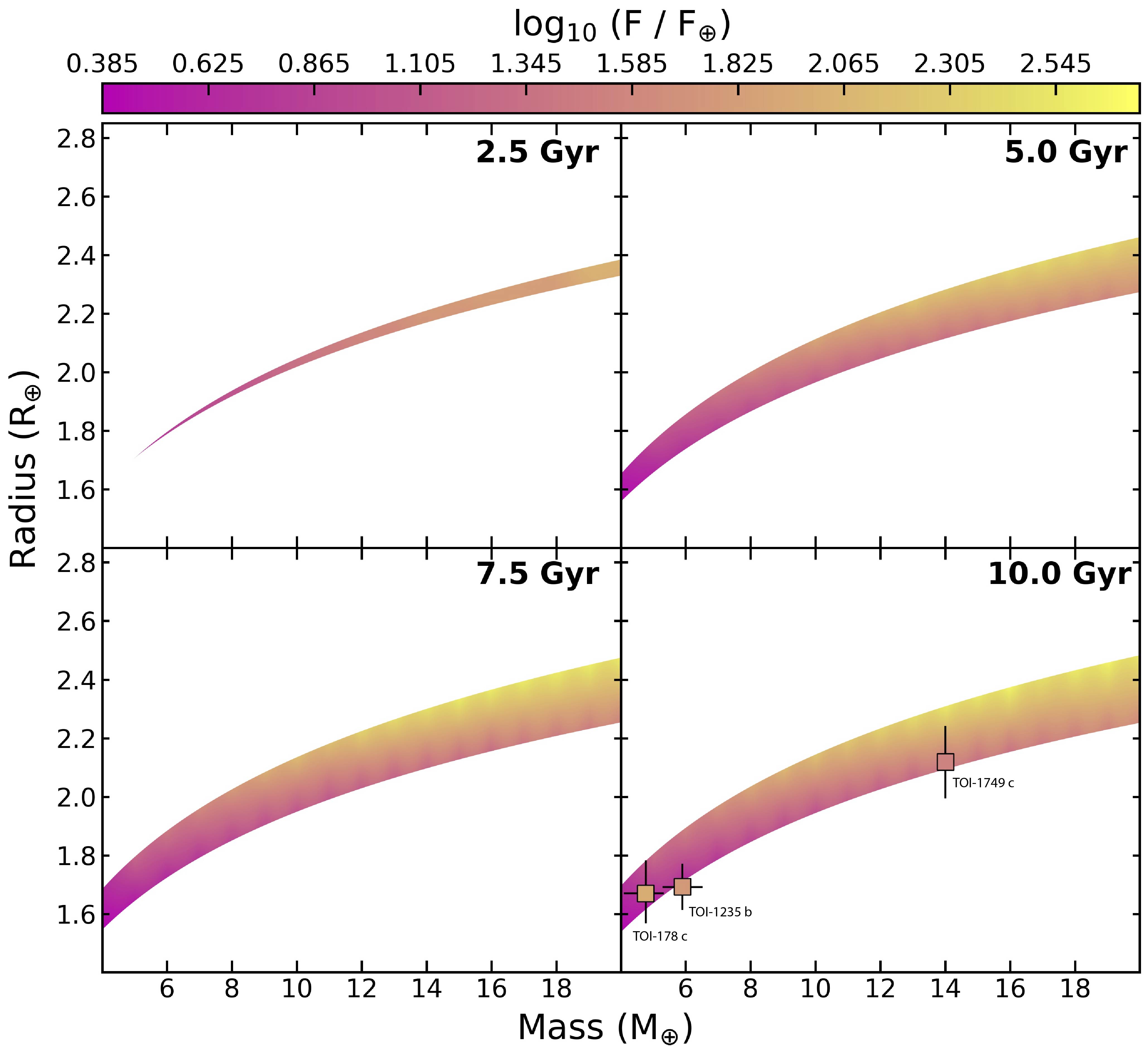}
    \caption{The mass-radius relationship for helium enhanced planets with homopause temperatures of 3,000 K at ages of 2.5, 5.0, 7.5, and 10.0 Gyr around a K type star. All model parameters other than host star types are identical to Figure \ref{fig:mr-3k-gstar}.}
    \label{fig:mr-3k-kstar}
\end{figure*}

\begin{figure*}[!htb]
    \includegraphics[width=1.0\linewidth]{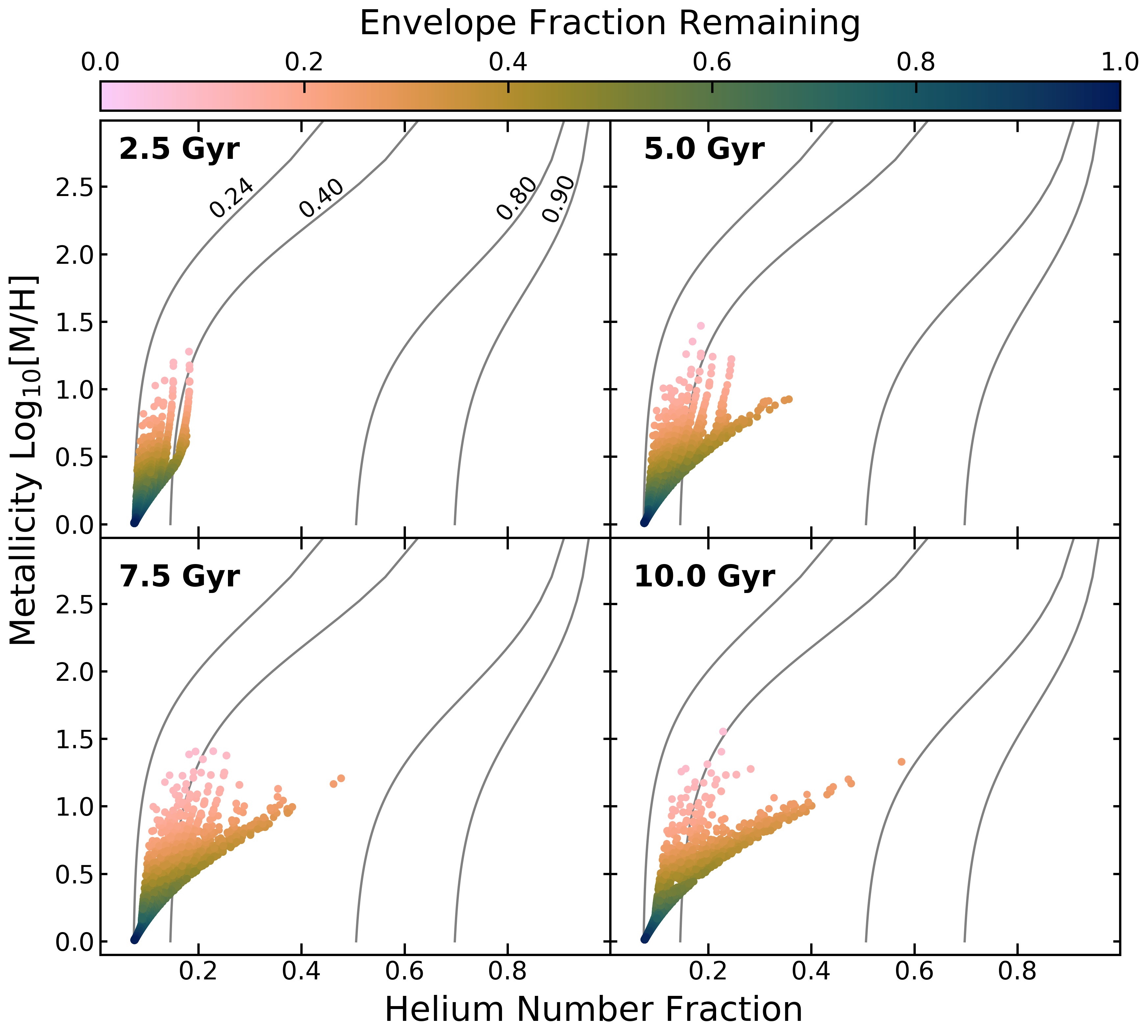}
    \caption{The metallicity and helium number fraction of planets with a homopause temperature of 3,000 K and evolved with fractionated mass loss at ages of 2.5, 5.0, 7.5, and 10.0 Gyr around a M type star. All model parameters other than host star types are identical to Figure \ref{fig:metals-3k-gstar}.}
    \label{fig:metals-3k-mstar}
\end{figure*}

\begin{figure*}[!htb]
    \includegraphics[width=1.0\linewidth]{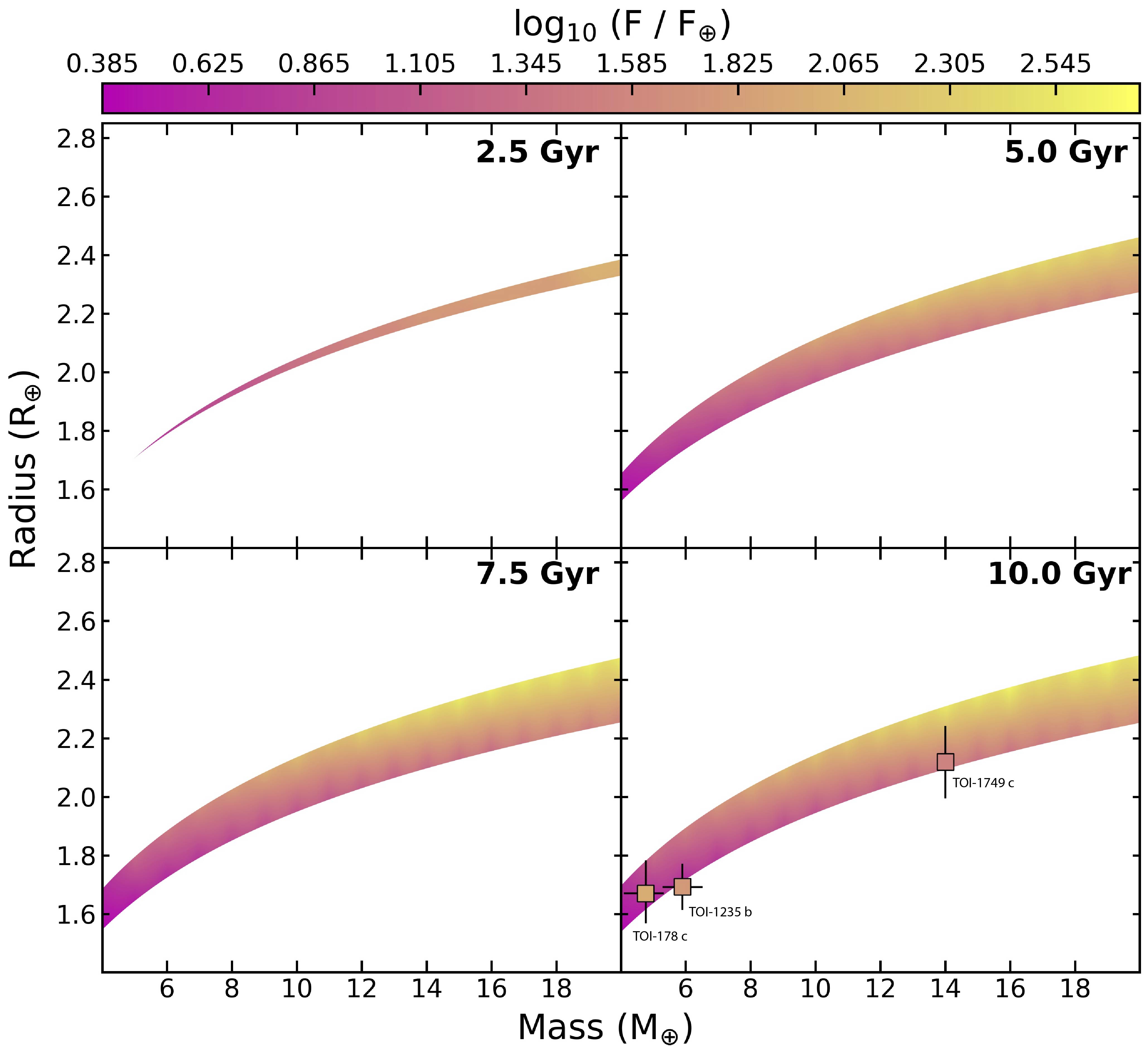}
    \caption{The mass-radius relationship for helium enhanced planets with homopause temperatures of 3,000 K at ages of 2.5, 5.0, 7.5, and 10.0 Gyr around a M type star. All model parameters other than host star types are identical to Figure \ref{fig:mr-3k-gstar}.}
    \label{fig:mr-3k-mstar}
\end{figure*}

Observations have shown that the location of the radius valley shifts to smaller radii for planets evolved around lower mass stars\cite{McDonald2019, Cloutier2020}. Figures \ref{fig:metals-3k-kstar}, \ref{fig:fr-3k-kstar}, \ref{fig:mr-3k-kstar}, \ref{fig:metals-3k-mstar}, \ref{fig:fr-3k-mstar}, and \ref{fig:mr-3k-mstar} show the $M_p-R_p-F_p$ parameter space of sub-Neptunes evolved with fractionated mass loss around K and M type stars. Compared to planets evolved around G type stars, planets orbiting cooler stars become helium enhanced at lower instellations, have smaller radii, and slightly less metal enhancement. For G stars we find the radius valley extended from approximately 1.6 R$_\oplus$ to 2.2 R$_\oplus$ and is approximately 0.2 R$_\oplus$ wide. In comparison, the radius valley for planets evolved around K and M type stars is narrower, as shown in Figures \ref{fig:fr-3k-mstar}, and \ref{fig:mr-3k-mstar}.

\subsection{Mass Loss Rates}\label{sec:mass_loss}
During the transonic escape regime, our models have escape rates of between 4$\times$10$^8$ to 2$\times10^9 \mathrm{g\,s^{-1}}$. Due to the translational and thermal energy losses, the escape rate remains nearly constant up to an age of approximately 1 Gyr. As the incident EUV decreases, the mass loss becomes energy limited and subsequently decreases approximately following a power law in time. By 10 Gyr the mass loss rates range from $\sim$ 5$\times$10$^6$ to 2$\times10^9 \mathrm{g\,s^{-1}}$. Therefore, planets with small (i.e., initial f$_{\rm env}$ $\leq$ 0.01) envelopes can lose the majority their primordial envelopes. For example, a 10.0 M$_{\oplus}$ planet with an initial envelope mass fraction of 0.5\% has an initial envelope mass of $\sim$3.0$\times$10$^{26}$~g. A sustained mass loss rate of 1$\times10^9~\mathrm{g\,s^{-1}}$ over 5.0 Gyr causes just over 50\% of the envelope mass to be lost.

Changing the mass loss efficiency factor has a large effect on the mass loss rate for sub-Neptunes. However, the mass loss efficiency factor is degenerate with orbital separation. Increasing the mass loss efficiency merely moves the parameter space in which planets become helium enhanced to larger orbital separations. Furthermore, the mass loss efficiency is not well constrained within the field\cite{MurrayClayEt2009ApJ, Owen&Wu2013ApJ, 2014A&A...571A..94S}. Throughout this work we assume a constant value of 10$\%$ following Malsky \& Rogers (2020)\cite{Malsky_2020}.

\subsection{Remnant Cores}
\begin{figure*}[!htb]
    \includegraphics[width=1.0\linewidth]{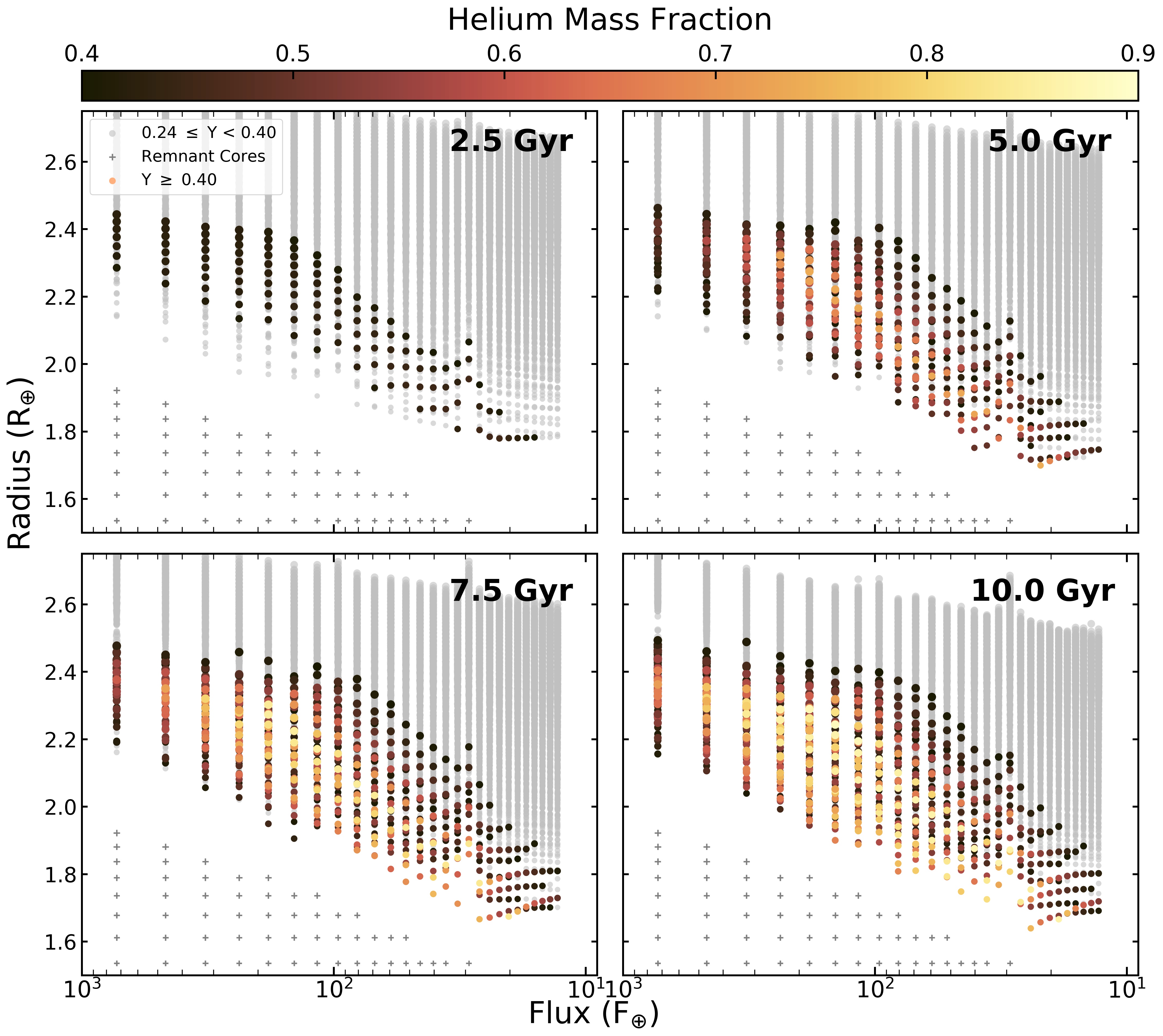}
    \caption{Simulated insolation planet flux-radius relations at ages of 2.5, 5.0, 7.5, and 10.0 Gyr. Planets have masses between 4.0 and 20.0 M$_\oplus$, orbital separations between 0.01 and 0.30 au, and initial envelope fractions between 0.001 and 0.01. All planets were evolved around a 6,000 K host star and began their evolutions with solar composition envelopes. The colored dots correspond to planets with helium mass fractions greater than 0.40, the gray dots correspond to planets with helium mass fractions between 0.24 and 0.40, and the gray pluses correspond to remnant cores which lost their entire envelopes before 2.5 Gyr. We consider planets with homopause (defined in \S~\ref{sec:more-methods}) temperatures of 3,000~K orbiting sun-like G stars, and in \S~\ref{sec:more-methods} show different homopause temperatures and host star spectral types affect helium enhancement.}
    \label{fig:fr-3k-gstar}
\end{figure*}

\begin{figure*}[!htb]
    \includegraphics[width=1.0\linewidth]{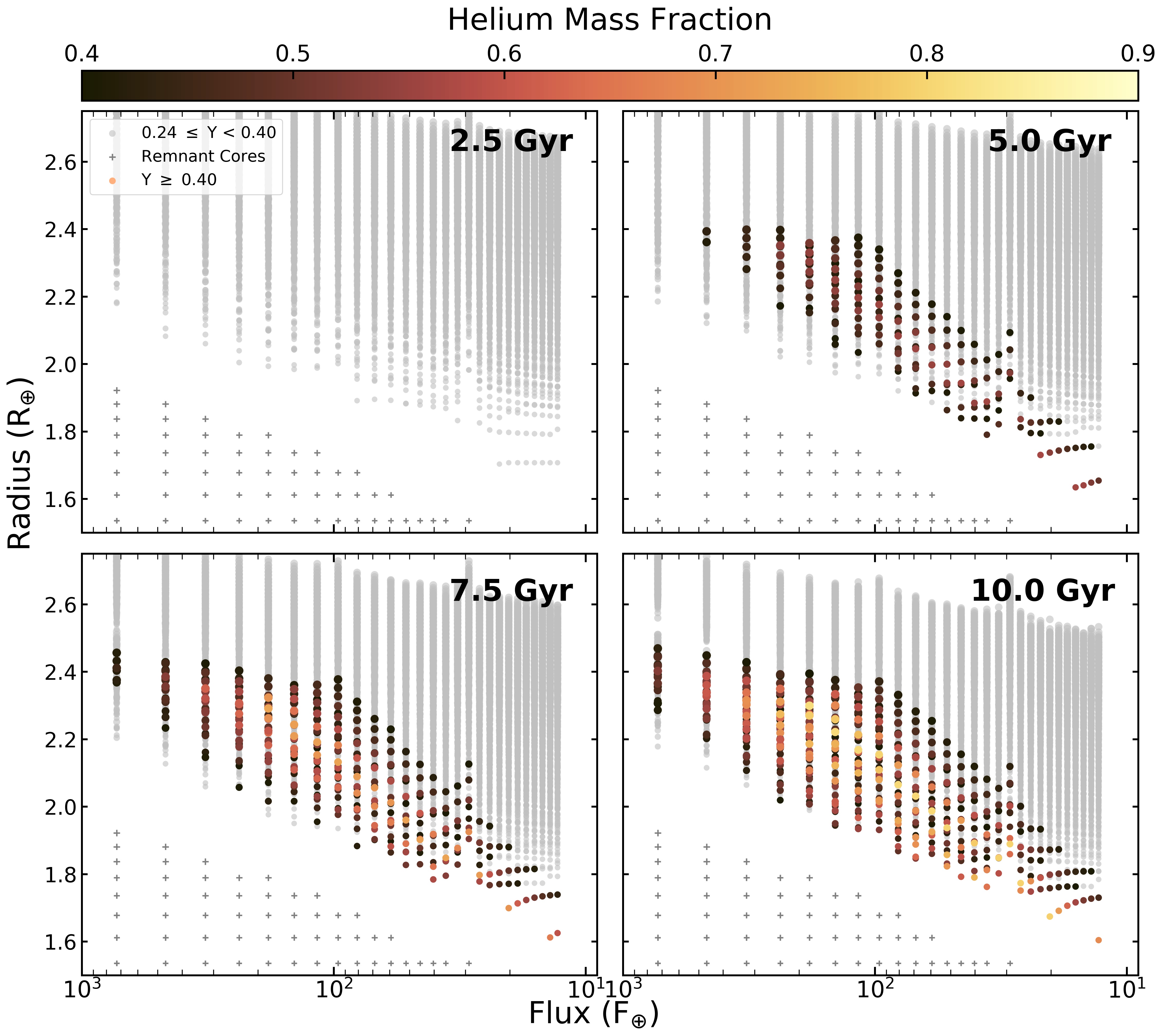}
    \caption{Planet flux-radius relation at ages of 2.5, 5.0, 7.5, and 10.0 Gyr for planets with a homopause temperature of 10,000 K and evolved orbiting a G type star. All model parameters other than homopause temperatures are identical to Figure \ref{fig:fr-3k-gstar}.}
    \label{fig:fr-10k-gstar}
\end{figure*}

\begin{figure*}[!htb]
    \includegraphics[width=1.0\linewidth]{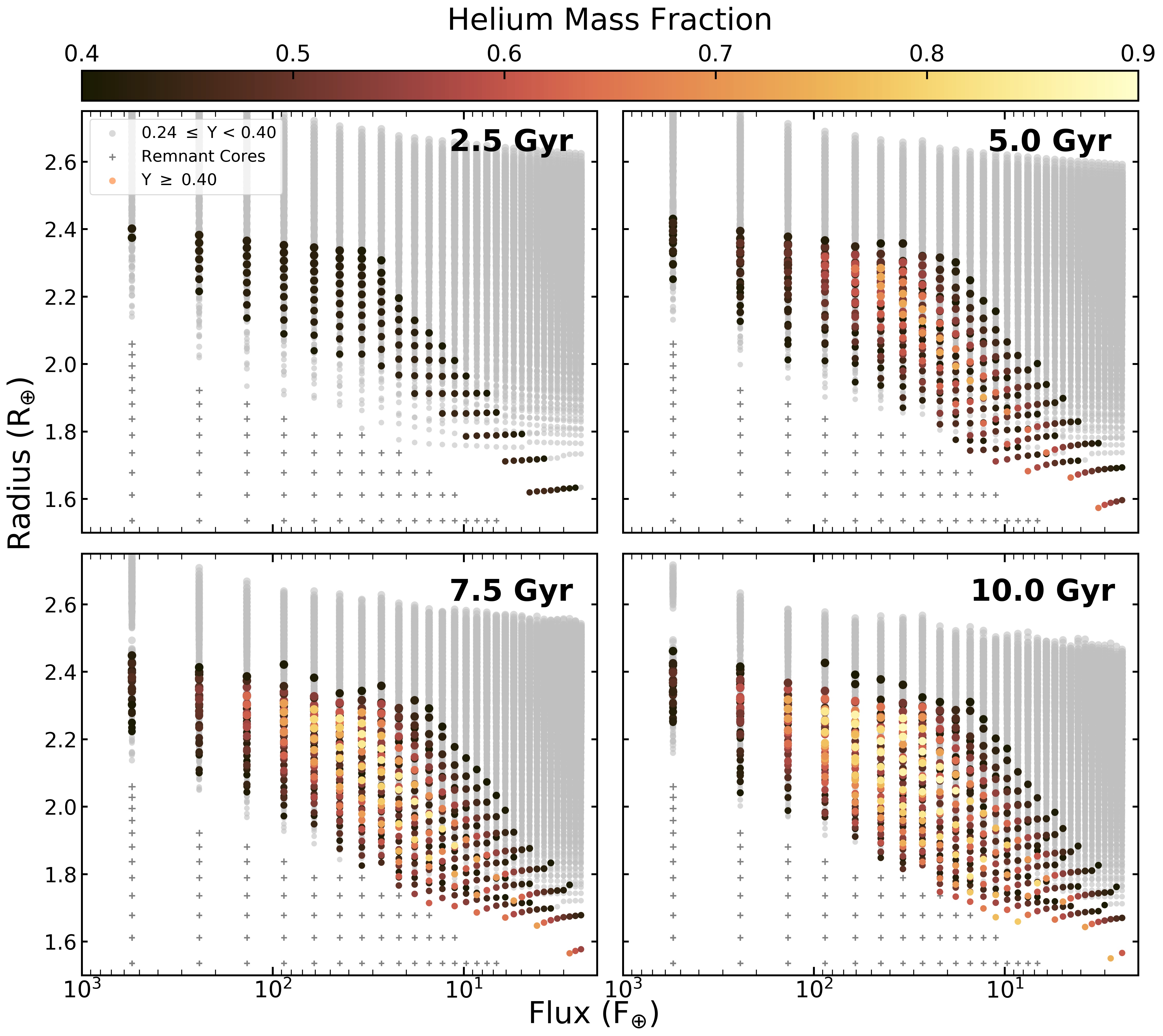}
    \caption{Planet flux-radius relation at ages of 2.5, 5.0, 7.5, and 10.0 Gyr for planets with a homopause temperature of 3,000 K and evolved orbiting a K type star. All model parameters other than host star types are identical to Figure \ref{fig:fr-3k-gstar}.}
    \label{fig:fr-3k-kstar}
\end{figure*}

\begin{figure*}[!htb]
    \includegraphics[width=1.0\linewidth]{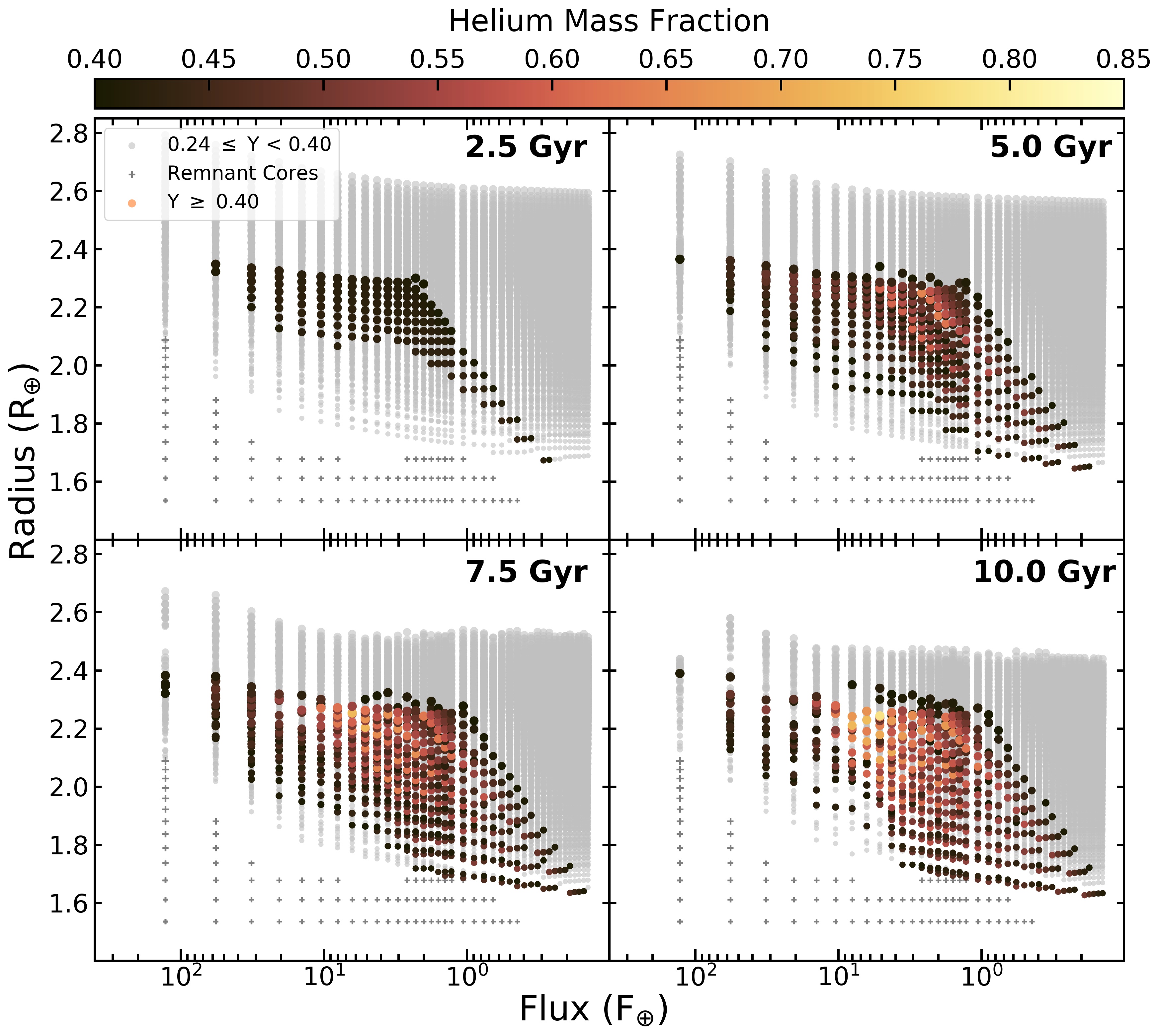}
    \caption{Planet flux-radius relation at ages of 2.5, 5.0, 7.5, and 10.0 Gyr for planets with a homopause temperature of 3,000 K and evolved orbiting a M type star. All model parameters other than host star types are identical to Figure \ref{fig:fr-3k-gstar}.}
    \label{fig:fr-3k-mstar}
\end{figure*}

In our simulations, a number of planets were stripped of nearly their entire envelope and failed to evolve for the full 10 Gyr that we simulated. We call these planets remnant cores and define them as any planet which failed to evolve past 2.5 Gyr. We assign these remnant cores radii equal to that of their rocky cores\citep{2011ApJ...738...59R}. We find remnant cores for fluxes between 14 and 1100 F$_\oplus$ for G stars, between 6 and 600 F$_\oplus$ for K stars, and between 0.4 and 110 F$_\oplus$ for M stars. When we compare the population of remnant cores to the helium enhanced planets, we see a clear bifurcation. Remnant cores have radii less than 2.2 R$_\oplus$, and occupy a $F_\oplus-R_\oplus$ parameter space below that of helium enhancement. Figures~\ref{fig:fr-3k-gstar}, \ref{fig:fr-3k-kstar}, and \ref{fig:fr-3k-mstar} show the population of helium enhanced planets, juxtaposed against the population of remnant cores. The largest rocky cores have radii that are equal to the smallest helium enhanced planets. These simulations were formed for a broad grid of initial conditions and we do not make any attempt to fine tune or match the empirical radius valley. The initial mass distribution of planets will affect the actual radius distribution of remnant cores/helium enhanced planets achieved.

\newpage
\clearpage
\bibliographystyle{naturemag}

\clearpage
\textbf{Acknowledgements}
IM would like to thank colleagues in his graduate program and Marianne Cowherd, who provided editorial suggestions on drafts of this manuscript. IM would also like to thank Nick Merhle, who provided code used to convolve the simulated spectra. IM acknowledges support from the Michigan Space Grant under Grant No. 80NSSC20M0124. LAR gratefully acknowledges support from NSF FY2016 AAG Solicitation 12-589 award number 1615089, and the Research Corporation for Science Advancement through a Cottrell Scholar Award. This research has made use of the NASA Exoplanet Archive, which is operated by the California Institute of Technology, under contract with the National Aeronautics and Space Administration under the Exoplanet Exploration Program. This work was also completed in part with resources provided by the University of Chicago’s Research Computing Center.

\end{document}